\documentclass[aip,apr,amsmath,amssymb,reprint]{revtex4-2}
\usepackage{hyperref}
\usepackage{graphicx}
\usepackage{dcolumn}
\usepackage{bm}

\usepackage[utf8]{inputenc}
\usepackage[T1]{fontenc}
\usepackage{mathptmx}
\usepackage{etoolbox}
\usepackage{upgreek}
\usepackage[arc-separator = \,]{siunitx}[=2]

\makeatletter
\def\@email#1#2{%
 \endgroup
 \patchcmd{\titleblock@produce}
  {\frontmatter@RRAPformat}
  {\frontmatter@RRAPformat{\produce@RRAP{*#1\href{mailto:#2}{#2}}}\frontmatter@RRAPformat}
  {}{}
}%
\makeatother
\begin{document}


\title{Quantum Sensing Using Atomic Clocks for Nuclear and Particle Physics}
\author{Akio Kawasaki}
\email{akio.kawasaki@aist.go.jp}
\affiliation{National Metrology Institute of Japan (NMIJ), National Institute of Advanced Industrial Science and Technology (AIST), 1-1-1 Umezono, Tsukuba, Ibaraki 305-8563, Japan}


\begin{abstract}
Technologies for manipulating single atoms have advanced drastically in the past decades. Due to their excellent controllability of internal states, atoms serve as one of the ideal platforms as quantum systems. One major research direction in atomic systems is the precise determination of physical quantities using atoms, which is included in the field of precision measurements. One of such precisely measured physical quantities is energy differences between two energy levels in atoms, which is symbolized by the remarkable fractional uncertainty of $10^{-18}$ or lower achieved in the state-of-the-art atomic clocks. Two-level systems in atoms are sensitive to various external fields and can, therefore, function as quantum sensors. The effect of these fields manifests as energy shifts in the two-level system. Traditionally, such shifts are induced by electric or magnetic fields, as recognized even before the advent of precision spectroscopy with lasers. With high-precision measurements, tiny energy shifts caused by hypothetical fields weakly coupled to ordinary matter or by small effects mediated by massive particles can be potentially detectable, which are conventionally dealt with in the field of nuclear and particle physics. In most cases, the atomic systems as quantum sensors have not been sensitive enough to detect such effects. Instead, experiments searching for these interactions have placed constraints on coupling constants, except in a few cases where effects are predicted by the Standard Model of particle physics. Nonetheless, measurements and searches for these effects in atomic systems have led to the emergence of a new field of physics. In some cases, they open new parameter spaces to explore in conventionally investigated topics, e.g., dark matter, fifth force, and other physics beyond the Standard Model. In other cases, these measurements provide alternative experimental approaches to established topics, e.g., variations of fundamental constants searched for astronomically and nuclear structure studied in high-energy scattering experiments. The use of atomic clocks as quantum sensors for phenomena originating from nuclear and particle physics evolved significantly in the past decades. This paper highlights the recent developments in the field.  
\end{abstract}

\maketitle

\section{Introduction to quantum sensing for nuclear and particle physics}
Quantum sensing is defined in the following three ways, as carefully described in Ref. \cite{RevModPhys.89.035002}. 

(i) a measurement of a physical quantity using quantized energy levels as a detector. 

(ii) a measurement of a physical quantity based on superposition of quantum states. 

(iii) a measurement of a physical quantity employing quantum entanglement to enhance its sensitivity. 

There are many physical realizations of quantized energy levels defined in (i). Examples are superconducting qubits \cite{AnnRevCondMatPhys.11.369,ApplPhysRev.6.021318}, mechanical harmonic oscillators including micro- or nanoparticles and suspended mirrors \cite{NatPhys.18.15,RepProgPhys.83.026401}, spin systems in solid state materials such as nitrogen-vacancy centers \cite{AnnRevPhysChem.65.83,PhysRep.528.1,RevModPhys.92.015004}, cold atoms \cite{Science.357.995} and molecules \cite{Science.357.1002} that are often trapped, and trapped ions \cite{ApplPhysRev.6.021314,Science.339.6124}. Among them, this paper focuses on cold atoms and ions. Laser cooling and trapping allows them to be kept at the temperature of $\sim1$ mK or lower without any cryogenic systems. Coherence can be maintained for seconds thanks to the excellent isolation from the environment. Rapidly developing trapping and manipulation techniques enable independent and arbitrary control of the quantum state of a single atom or ion. This capability makes these systems suitable for quantum sensors. 

\begin{figure}
\centering
	\includegraphics[bb=0 0 336 299,width=0.5\linewidth]{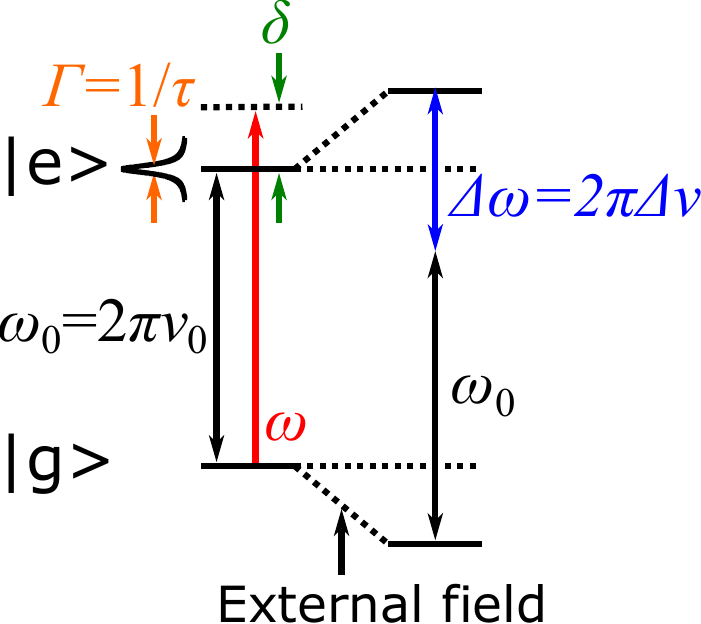}
\caption{A two-level system and relevant quantities. $|g \rangle$: the ground state. $|e \rangle$: the excited state. $\hbar \omega_0=h \nu_0$: the energy difference between these two states. $\omega$: frequency for the oscillating electromagnetic field, with $\delta=\omega-\omega_0$. $\Gamma=1/\tau$: natural linewidth of the excited state, with $\tau$ being its lifetime. $\Delta \omega=2 \pi \Delta \nu$: shift in resonant frequency induced by an external field.}
	\label{FigTwoLevel}
\end{figure}

When an atomic system is selected as a quantum system, it generally has many energy levels. For quantum sensing purpose, specific energy levels are typically selected, and control of the states is performed by applying oscillating electromagnetic fields on or near resonance. In such cases, the system can be well approximated as a two-level system consisting of the ground state $|g \rangle$ and the excited state $|e \rangle$. The relevant quantities are summarized in Fig. \ref{FigTwoLevel}. The energy difference between the two energy levels is $\hbar \omega_0=h\nu_0$, where $h=2 \pi \hbar$ is the Planck constant. The excited state has a natural decay linewidth denoted by $\Gamma$. The oscillating electromagnetic field has a frequency of $\omega$ detuned by $\delta$ from $\omega_0$. 

An external field can induce a frequency shift of $\Delta \omega$ from $\omega_0$ in the two-level system. $\Delta \omega$ can be converted to the magnitude of the external field, with a known relation between the field strength and $\Delta \omega$. One example is the frequency shift caused by a magnetic field, which is known as the Zeeman effect. When a small external magnetic field $B$ is applied, assuming that the ground state is insensitive to the magnetic field, $\Delta\omega=\mu_B g_F m_F B$, where $\mu_B$ is the Bohr magneton, $g_F$ is the Land\'e g-factor for the excited state, and $m_F$ is the magnetic quantum number for the total angular momentum $F$. Frequency shifts can be caused by some other fields. Another well-known effect is the Stark effect, which is induced by an external electric field. 

The external field is not necessarily limited to electromagnetic or other well-established fields in conventional physics. Consider a hypothetical bosonic field $\phi$ that induces $\Delta \omega$, whose constituent particle has a mass and thus can be a candidate for dark matter (such fields are discussed in Section \ref{SecUltralightDM}). If a two-level system is sufficiently sensitive, the $\phi$ field can be detected using this system. Searching for such a hypothetical field effectively serves as a search for new particles. Even if no frequency shift is observed, the measurement sets an upper bound on the interaction strength between the $\phi$ field and the two-level system, meaning that the interaction is weaker than the sensitivity of the two-level system.  This falls within the realm of physics beyond Standard Model (BSM), which is one of the major topic in contemporary particle physics. Another example is the sensitivity of atomic spectroscopy to nuclear charge distribution. The finite size of the nucleus causes a deviation in the energy levels of the electronic states in an atom compared to the case of a point charge. As a result, slight differences in nuclear charge radii between isotopes can be detected as shifts in transition frequencies between isotopes for the same electronic transition, which is called the isotope shift. Here, a structure of $\sim 1$ fm, which corresponds to an energy scale of $\gtrsim 1$ MeV, is probed by spectroscopy of electronic excitations at an energy scale of $\sim 1$ eV. 

These two examples illustrate how the quantum sensors, particularly those with precision spectroscopy, are sensitive to the nuclear and particle physics. One approach is to try to detect hypothetical interactions with small couplings to the Standard Model (SM) particles through high-precision measurement, such as state-of-the-art atomic clocks. Examples covered in this paper include searches for time variation of fundamental constants, ultralight dark matter, and new force between an electron and a neutron. Although axions and axion-like particles fall within this type of interaction, they are not discussed in this paper. Readers are encouraged to refer to some other review papers \cite{RevModPhys.97.025005,AnnRevNuclPartSci.65.485,AnnRevNuclPartSci.71.225,PhysRep.870.1} that deal with them in detail. Theoretical proposals for these interactions have been tested experimentally, and so far, the experiments give null results, setting constraints on such hypothetical interactions. Another approach involves phenomena conventionally studied through high-energy collisions. In this case, some precision measurement experiments detect high-energy phenomena in atomic systems. The topics covered in this paper are nuclear structure determination via isotope shifts. 

The effects of nuclear and particle physics phenomena are typically extremely small, compared to the conventional shifts caused by external electromagnetic fields. Distinguishing these tiny shifts from those induced by conventional electromagnetic fields is crucial. In this paper, the term "signal" refers to the quantity aimed to be detected: namely, the effects originating from nuclear and particle physics, e.g., the effect of the $\phi$ field. All other effects that behave similarly to the targeted signal are regarded as "backgrounds". These backgrounds include shifts caused by conventional external fields such as electric and magnetic fields. Even among effects induced by nuclear and particle physics, one can be considered as the signal, while others act as the backgrounds. For example, in searches for BSM physics, other effects from the SM are backgrounds. To discriminate the signal from backgrounds properly, multiple methods are employed. In addition, a proper theoretical framework is essential to extract nuclear and particle physics effects from slight energy shifts. The following discussions focus on how to select an appropriate two-level system to extract a desired nuclear and particle physics phenomenon, and how to cancel all backgrounds that can potentially mimic the signal. 

Definition (ii) is also broad. In a two-level system, a superposition state $(|g \rangle + | e \rangle)/\sqrt{2}$ is frequently used to detect a phase drift between the two-level system and the driving oscillating electromagnetic field (see Section \ref{SecAMO}). Among the three definitions of quantum sensing, Definition (iii) is the most stringent. Entangled states, such as squeezed states and Schr\"odinger's cat states, are typically difficult to generate and maintain. As a result, examples of measurements with entangled states are limited. This paper discusses quantum sensing with entanglement when relevant reports are available. Experimental results are covered in Section \ref{SecAtomicClock}.

This paper first briefly summarizes how atomic systems are prepared as quantum sensors. This section is intended for readers outside of the field of atomic, molecular, and optical physics (AMO) to offer a glimpse of how AMO experiments operate. As a realization of the two-level system, optical atomic clocks are first discussed. This is partly because precision spectroscopy of atomic two-level systems forms the foundation of precise determination of the energy difference in the two-level system and because optical atomic clocks are the state-of-the-art in this field. Additionally, atomic clocks serve as a platform for exploring nuclear and particle physics through precision measurements. The systematic shifts and uncertainties discussed here provide typical factors to consider in other types of precision spectroscopy. This part includes details of AMO physics. The discussion is followed by searches for variation of fundamental constants. This shows the sensitivity of atomic clocks as quantum sensors, also introducing the importance of two clock architectures: highly charged ion (HCI) clocks and nuclear clocks. These two systems are not only promising candidate for next-generation optical clocks \cite{PhysRevA.86.022517,EurophysLett.61.181}, but also serve as quantum sensors for variations of fundamental constants and other nuclear and particle physics phenomena. A related topic is searches for ultralight dark matter, where periodic modulation of fundamental constants is interpreted as potential signals of ultralight dark matter. After this, searches for a new force between an electron and a neutron through isotope shifts are discussed, which also function as precise measurements of the nuclear structure. The section for the variation of fundamental constants, ultralight dark matter, and fifth force search with isotope shifts are structured with separate discussions on the theoretical frameworks and experimental results. These subsections provide overviews of nuclear and particle physics theories and AMO experiments, respectively. 

How are these topics related to quantum sensors? As introduced, the basis of quantum sensors discussed in this paper is the two-level system. In an atomic clock, a two-level system is selected as the source of stability. Various systematic shifts, mainly induced by external electromagnetic fields, are suppressed as a clock. This opens the possibility of detecting small signal originating from nuclear and particle physics phenomena. Examples of such signals are the variation of fundamental constants, ultralight dark matter, a new force between an electron and a neutron, and nuclear charge distribution. 

\section{Typical procedure to utilize atoms for quantum sensors}\label{SecAMO}
In this section, typical experimental procedures for preparing atoms as quantum sensors are briefly summarized. In some cases, for example, ion trap experiments, certain cooling and trapping procedures are different from those described below for neutral atoms. 

Trapping atoms begins with generating an atomic beam, a process that depends on the vapor pressure of the atoms. For low-vapor-pressure atoms such as Yb or Sr, an atomic oven heated to a few hundred degrees Celsius is used. For alkali atoms, which have relatively high vapor pressure, atomic dispensers are commonly used. The resulting atomic beam has a center velocity of a few hundred m/s, requiring deceleration for trapping. A Zeeman slower \cite{PhysRevLett.48.596} is sometimes employed for this purpose, where a circularly polarized laser beam slightly off-resonant from a broad atomic transition propagates to the opposite direction of the atomic beam and a spatially varying magnetic field is also applied. This combination slows down atoms within a specific velocity range to $O(10)$ m/s. 

Sufficiently slow atoms can be trapped by a magneto-optical trap (MOT) \cite{PhysRevLett.59.2631}. A MOT is formed by a combination of circularly polarized light and a quadrupole magnetic field. In the standard configuration, counter-propagating laser beams with $\sigma^+/\sigma^-$ polarization, red-detuned from a broad-linewidth transition, provide a dissipation term. The quadrupole magnetic field induces a position-dependent scattering rate, generating an approximately harmonic potential.  As a result, the atoms are simultaneously slowed, cooled, and confined within the trap. For a transition with $\Gamma=2\pi \times 5$ MHz, the Doppler-limited temperature is $\hbar \Gamma/2k_B=0.12$ mK, where $k_B$ is the Boltzmann constant. Using a narrower-linewidth transition, e.g., $\Gamma=2\pi \times 50$ kHz, atoms can be cooled down to $\sim 1$ $\upmu$K. In addition, for atoms with Zeeman sublevels in the ground state, polarization gradient cooling \cite{JOSAB.6.2023} can further reduce the temperature below the Doppler limit.

After being trapped in a MOT, atoms are typically transferred to an optical lattice. An optical lattice is formed by pairs of counter-propagating laser beams that create standing waves. Due to the ac Stark shift, atoms are attracted to the most (least) intense part of the light field when the trapping laser is red- (blue-) detuned from broad-linewidth transitions \cite{AdvAtomMolOptPhys.42.95}. Depending on the number and configuration of the pairs of the counterpropagating laser beams, a one-, two-, or three-dimensional lattice can be formed. 

For precision spectroscopy, the key role of an optical lattice is to put atoms in a recoil-free environment. Thermal atoms that are not trapped change their motional states upon absorbing or emitting a photon. The energy changes due to this recoil is called recoil energy $E_R=\hbar \omega_{\rm rec}=h^2 c^2/2m_{\rm atom}\nu^2$, where $m_{\rm atom}$ is the mass of the atom, $c$ is the speed of light, and $\omega_{\rm rec}$ is the recoil frequency. When the trapping frequency of atoms along the direction of the probe light resonant with the two-level system is $\omega_{\rm trap}$, and $\omega_{\rm rec}/\omega_{\rm trap} =\eta^2 \ll 1$, the atoms remain in their motional states after absorbing or emitting a photon, making the spectroscopy recoil-free. Intuitively, the momentum of the photon is absorbed by the entire trap, similar to the M\"ossbauer effect. $\eta^2$ is called the Lamb-Dicke parameter, and when the condition $\eta^2 (2n_z+1) \ll1$ is satisfied, where $n_z$ is the quantum number for for the longitudinal vibrational state, atoms are in Lamb-Dicke regime, where the atoms do not change their motional state when they absorb or emit photons. 

In an optical lattice, atoms can be further cooled using various techniques. Demonstrated methods include evaporative cooling \cite{AdvAtomMolOptPhys.37.181}, sideband-resolved Raman cooling \cite{PhysRevLett.75.4011,PhysRevLett.80.4149}, cooling with a narrow-linewidth transition \cite{PhysRevLett.129.113202}, and Sisyphus cooling \cite{PhysRevLett.133.053401}. The final temperature can be very low, sometimes 100 nK or lower. Colder atoms with lower average energy are essential for achieving trapping in a shallow optical lattice, typically used in the state-of-the-art optical lattice clocks. 

The final stage of state preparation involves initializing all atoms into a single internal state. The ground state of an atom generally consists of multiple sublevels, such as Zeeman sublevels and hyperfine structures. In general, atoms are distributed over these states, but to utilize them as quantum sensors, they need to be initialized in a single state. Optical pumping \cite{PhysRev.91.1008,RevModPhys.44.169} is the simplest method for this purpose. The exact procedure depends on the system. For example, if the ground state has several Zeeman sublevels with different $m_F$ but no hyperfine structure, irradiating the atoms with $\sigma^+$-polarized light resonant with a broad-linewidth transition can transfer all atoms into the highest $m_F$ state. More sophisticated techniques such as stimulated Raman adiabatic passage\cite{RevModPhys.89.015006,JChemPhys.92.5363}, rapid adiabatic passage \cite{PhysLettA.29.369,PhysRevLett.32.814}, and coherent population trapping \cite{OpticaActa.32.61} are also widely used for state preparation. 

After state preparation is completed, atoms undergo a measurement sequence. While the measurement sequence depends on experiments, there are two basic procedures for a sequence. The sequence is performed on a two-level system as illustrated in Fig. \ref{FigTwoLevel}. The laser intensity is characterized by the Rabi frequency $\omega_R$. A more detailed formalism can be found in standard textbooks on atomic physics \cite{API}. 

The first sequence is the Rabi sequence, where resonant light drives the two-level system. If the initial state is $|g \rangle$, the final state $|f \rangle$ after a pulse of duration $t$ satisfies $|\langle e|f \rangle|^2 = \frac{\omega_R^2}{\omega_R^2+\delta^2}\sin^2\frac{\sqrt{\omega_R^2+\delta^2}t}{2}$. When $\delta=0$ and $\omega_R t=\pi$, $|f \rangle = |e \rangle$. Such a pulse to drive the system from $|g \rangle$ to $|e \rangle$ is called a $\pi$ pulse. If the pulse duration is half of the $\pi$ pulse, $|f \rangle=(|g \rangle + | e \rangle)/\sqrt{2}$, and this pulse is referred to as a $\pi/2$ pulse. The relative phase between $|g \rangle$ and $|e \rangle$ in $|f \rangle$ can be controlled by adjusting the phase offset of the driving light field. If $\omega_0$ is perturbed by external fields, the nominal pulse duration may not result in the desired final state. To mitigate this, composite pulses, which are extensively studied in the field of nuclear magnetic resonance \cite{JMagnReson.48.234,RevModPhys.76.1037}, can be applied in laser spectroscopy. 

The second basic sequence is the Ramsey sequence. In this sequence, atoms are initially prepared in $|g \rangle$. First, a $\pi/2$ pulse is applied, placing the atoms in a superposition state $(|g \rangle + | e \rangle)/\sqrt{2}$. Atoms are then left for a duration of $\tau_R$, which is referred to as the interrogation time. During this period, if the laser frequency $\omega(t)$ generally dependent on time $t$ is different from $\omega_0$, the accumulated phase $\Delta \phi = \int_0^{\tau_R} (\omega(t)-\omega_0)dt$ is recorded as a phase shift between the atomic system and the laser. $\Delta \phi$ is read out by another $\pi/2$ pulse and subsequent proper state detection of $|g \rangle$ and $|e \rangle$. Note that when $\omega(t)\simeq\omega$ is approximately constant, $\Delta \phi /\tau_R \simeq \delta $, meaning that the Ramsey sequence can be used to measure $\delta$. If $\delta=0$ and $\omega_0$ is shifted by $\Delta \omega$ due to an external field, $\Delta \phi/\tau_R=-\Delta \omega$ serves as a sensor for $\Delta \omega$. Both the Rabi and Ramsey sequences are used to measure the frequency offset between the laser system and the atomic system to lock the laser to the atomic resonance. 

To function as a sensor, the final state needs to be detected. The most common detection method involves driving a broad-linewidth transition, e.g., $\Gamma/2\pi \gtrsim 1$ MHz, from $|g \rangle$ using a light field and measuring the response of the atoms. One method is fluorescence imaging, where scattered photons from the atoms are collected by a photodetector or a camera. Alternatively, absorption imaging can be used, in which a camera records the probe light transmitted through the atomic cloud. The presence of atoms causes attenuation of the probe beam due to light scattering in other directions, resulting in a shadow in the image. To obtain the fraction of atoms in $|g \rangle$ over the total atom number, population in $|e \rangle$ also needs to be detected. 
This is achieved by transferring atoms from $| e \rangle$ back to $|g \rangle$ and imaging them separately, after removing all atoms originally in $|g \rangle$. 

\section{Atomic clock}\label{SecAtomicClock}
\subsection{Introduction}
\subsubsection{Components of clocks}
Modern stable clocks, including atomic clocks, consist of three main components: an oscillator, a counter, and a reference. The periodic motion of the oscillator is counted by the counter, which displays the time. For the stability purpose, the oscillator is locked to the reference. Nowadays, the most stable and accurate clocks are realized by optical atomic clocks, i.e., an atomic clock using an optical transition for clock operation. In an optical atomic clock, these three components are physically realized as a stable laser, a frequency comb, and an atomic transition in or near optical frequencies, respectively. The frequency of the stable laser is tuned to the resonant frequency of the atomic transition, so that the response of atoms as a result of interrogation by the laser can serve as a signal for feedback to stabilize the laser frequency. Since the atomic transition provides the ultimate source of stability, a deep understanding of the behavior of atoms is crucial for improving the accuracy of these clocks. 

When a system is regarded as stable, measurements performed in the system at different times yields the same value. The variation of the measured value is characterized by uncertainty, which is categorized into two types: statistical uncertainty and systematic uncertainty. Statistical uncertainty reflects the spread of measured values when repeated measurements are performed under the same conditions. Systematic uncertainties accounts for uncertainties other than the statistical uncertainty. In atomic clocks, systematic uncertainties typically include the uncertainties on the estimated frequency shifts induced by external fields and other factors. More stable systems enable higher precision measurements, because of their smaller statistical uncertainty. Precision is defined by a small fractional uncertainty, i.e., the uncertainty divided by the measured value. Accuracy refers to how close the measurement value is to the true value. 

For a stable reference, it is desirable to have a high sensitivity to small frequency shifts. This requirement can be achieved by using a transition with a large quality factor $\omega_0/\Gamma$. Such transitions are forbidden transitions, where a forbidden transition is defined as a transition that cannot be driven by the electric dipole (E1) interaction. The absence of transition probability due to the strongest interaction between atoms and light results in a long-lived excited state, if the excited state is without an alternative decay path to the ground state allowed by E1 transitions. In addition, for atomic clocks, it is crucial to select an atomic resonance with minimal sensitivity to external fields. 

\begin{figure}
\centering
	\includegraphics[bb=0 0 567 378,width=1\linewidth]{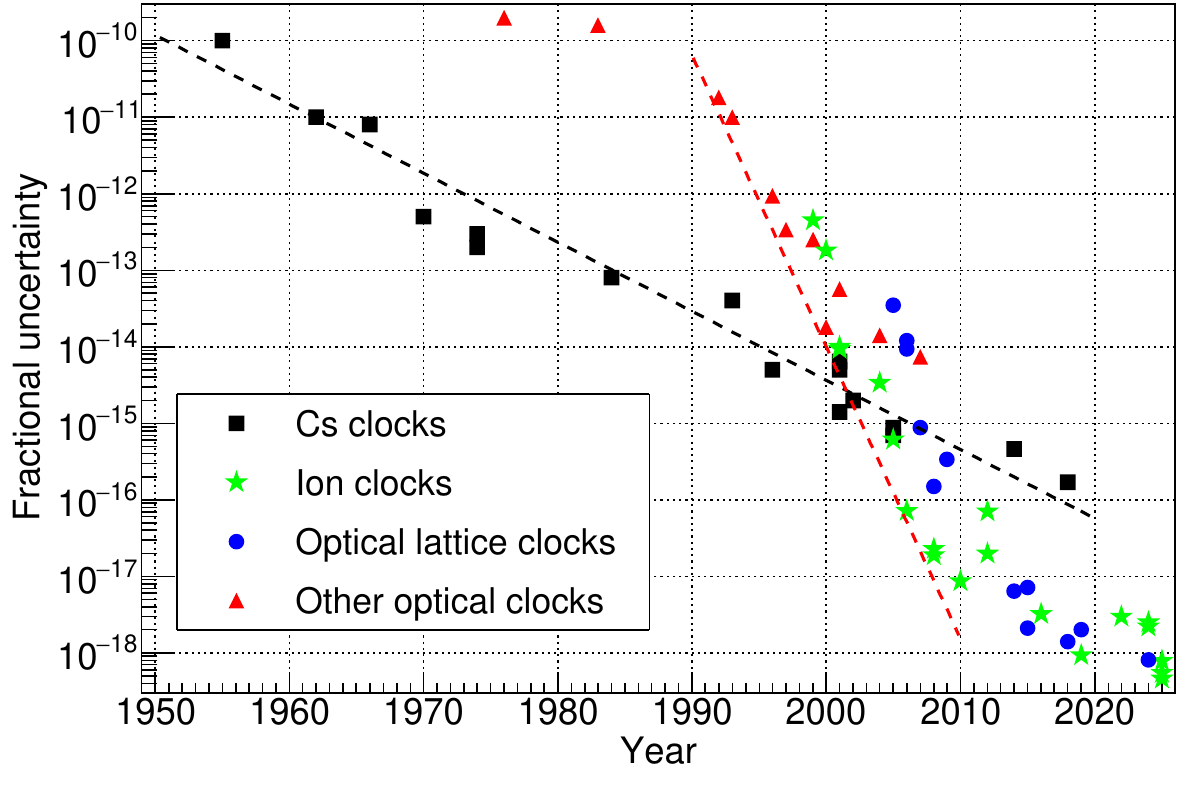}
\caption{Historical development of the fractional uncertainty of atomic clocks: the black dotted line shows the average improvement in the accuracy for the Cs clocks, and the red dotted line indicates approximate development of the accuracy for the best optical atomic clocks. The data points correspond to total fractional uncertainties reported in the following references. 
Cs clocks: Refs.
\cite{PhilTransRoySocLondon.250.45,IRETrans.11.231,IEEETransInstrumMeas.15.48,IEEETransInstrumMeas.19.156,IEEETransInstrumMeas.23.489,NBSSpecPub.617.25,IEEEIntFreqCtrlSymp.71,Proc28thAPTTISAM.225,Metrologia.38.427,Metrologia.39.321,Metrologia.42.411,Metrologia.51.174,Metrologia.55.789,Metrologia.38.343,JPhysB.38.S449,PhysRevLett.82.4619}
Ion clocks: Refs.
\cite{PhysRevLett.82.3228,OptLett.25.1729,PhysRevLett.86.4996,OptLett.26.1589,Science.306.1355,PhysRevLett.94.230801,PhysRevLett.97.020801,Science.319.1808,PhysRevLett.104.070802,PhysRevLett.109.203002,PhysRevLett.116.063001,PhysRevLett.123.033201,Metrologia.61.045001,PhysRevLett.135.033201,PhysRevAppl.24.044082,2506.17423}
Optical lattice clocks: Refs.
\cite{Nature.435.321,PhysRevLett.97.130801,JPhysSocJpn.75.104302,PhysRevLett.98.083002,Science.319.1805,PhysRevLett.103.063001,Nature.506.71,NatPhoton.9.185,NatCommun.6.6896,Nature.564.87,Metrologia.56.065004,PhysRevLett.133.023401,PhysRevLett.134.023201}
Other optical atomic clocks: Refs.
\cite{ApplOpt.15.734,OptLett.8.136,PhysRevA.35.4878, PhysRevA.39.4591,PhysRevLett.69.1923,OptCommun.97.29,PhysRevLett.76.18,PhysRevLett.79.2646,IEEETransInstrumMeas.48.613,PhysRevLett.84.5496,PhysRevLett.86.4996,PhysRevLett.92.230802,Metrologia.44.146}. 
}
	\label{FigClockStability}
\end{figure}

Historically, microwave transitions between $m_F=0$ states of two hyperfine levels in the ground state of alkali atoms, such as Cs and Rb, were used as the atomic transitions. Over the decades, the accuracy of the Cs atomic clock has improved by six orders of magnitude, as shown in Fig. \ref{FigClockStability}. With the advent of laser spectroscopy, optical atomic clocks have advanced much more rapidly than microwave atomic clocks. One advantage of optical atomic clocks is the transition frequency $\nu_0$, which is five orders of magnitude larger than that for microwave atomic clocks and thus helps increase $\omega_0/\Gamma$. 

Optical atomic clocks are sorted into two types: optical lattice clocks and ion clocks. An optical lattice clock interrogates thousands of atoms trapped in an optical lattice, whereas an ion clock conventionally operates with a single ion trapped in a Paul trap. Initially, ion clocks were the leading optical atomic clocks in terms of performance. The optical lattice clock was proposed in 2001 \cite{FreqStdMetrol.323,PhysRevLett.91.173005}, and first demonstrated in 2005 \cite{Nature.435.321}. Around this time, the accuracy of optical atomic clocks started to surpass that of microwave atomic clocks, and since then, their accuracies improved by three orders of magnitude. Currently, the smallest fractional uncertainties achieved by optical lattice clocks and ion clocks are comparable, with each type offering distinct advantages and disadvantages. The early development of optical atomic clocks has been summarized well in previous review papers \cite{,LRDNC.036.555,RevModPhys.87.637}. 

\subsubsection{Allan deviation and the stability of atomic clocks}
The stability of atomic clocks is typically described by the Allan deviation $\sigma(T)$ \cite{ProcIEEE.54.221}. It is a widely used measure of the stability of time-varying signals and is defined as follows:
\begin{equation}
\sigma^2(T)=\frac{1}{2}\langle \left( \bar{y}_{i+1}-\bar{y}_i \right)^2 \rangle 
=\frac{1}{2T^2}\langle \left( x_{i+2}-2x_{i+1}+x_i \right)^2 \rangle,
\end{equation}
where $\langle \cdots \rangle$ denotes the expectation operator, $\bar{y}_i=(x(iT_C+\tau)-x(iT_C))/T$, and $x_i=x(iT_C)$ is the $i$th data point taken at time intervals of $T_C$. The Allan deviation follows a power-law behavior depending on the dominant noise source. It scales as $T^{-1}$, $T^{-1/2}$, $T^{0}$, $T^{1/2}$, and $T$ for the phase noise, flicker frequency noise, white frequency noise, random walk frequency noise, and frequency drift, respectively.


The stability of state-of-the-art atomic clocks described by the Allan deviation is limited by quantum mechanical randomness when a superposition state between $|g \rangle$ and $|e \rangle$ collapses into either $|g \rangle$ or $|e \rangle$ upon measurement. This effect is known as quantum projection noise (QPN) \cite{PhysRevA.47.3554}, and the corresponding stability limit is called the standard quantum limit (SQL). The SQL is given by 
\begin{equation}
\sigma_{T}=\frac{1}{\omega_0 \tau_R}\sqrt{\frac{T_C}{T}}\sqrt{\frac{\xi_W^2}{N}},
\end{equation}
where $T_C$ is the length of a single measurement cycle, $T$ is the total measurement time, $\xi_W^2$ is the Wineland parameter for spin squeezing \cite{PhysRevA.50.67}, and $N$ is the number of atoms in the system \cite{Nature.588.414}. For a coherent spin state \cite{PhysRevA.6.2211}, $\xi_W^2=1$, while $\xi_W^2<1$ when the spin state is squeezed in the phase direction during the Ramsey sequence.

\subsection{Popular atomic species for atomic clocks}
For ion clocks and optical lattice clocks, atoms with certain kinds of transitions are preferred. A narrow-linewidth transition is essential for clock operation. Also, a broad-linewidth transition is important for efficient laser cooling. Atoms with both types of transitions can be sorted into two main groups.

\subsubsection{Alkali-atom-like species}\label{SecAAtoms}
\begin{figure}
\centering
	\includegraphics[bb=0 0 683 349,width=1\linewidth]{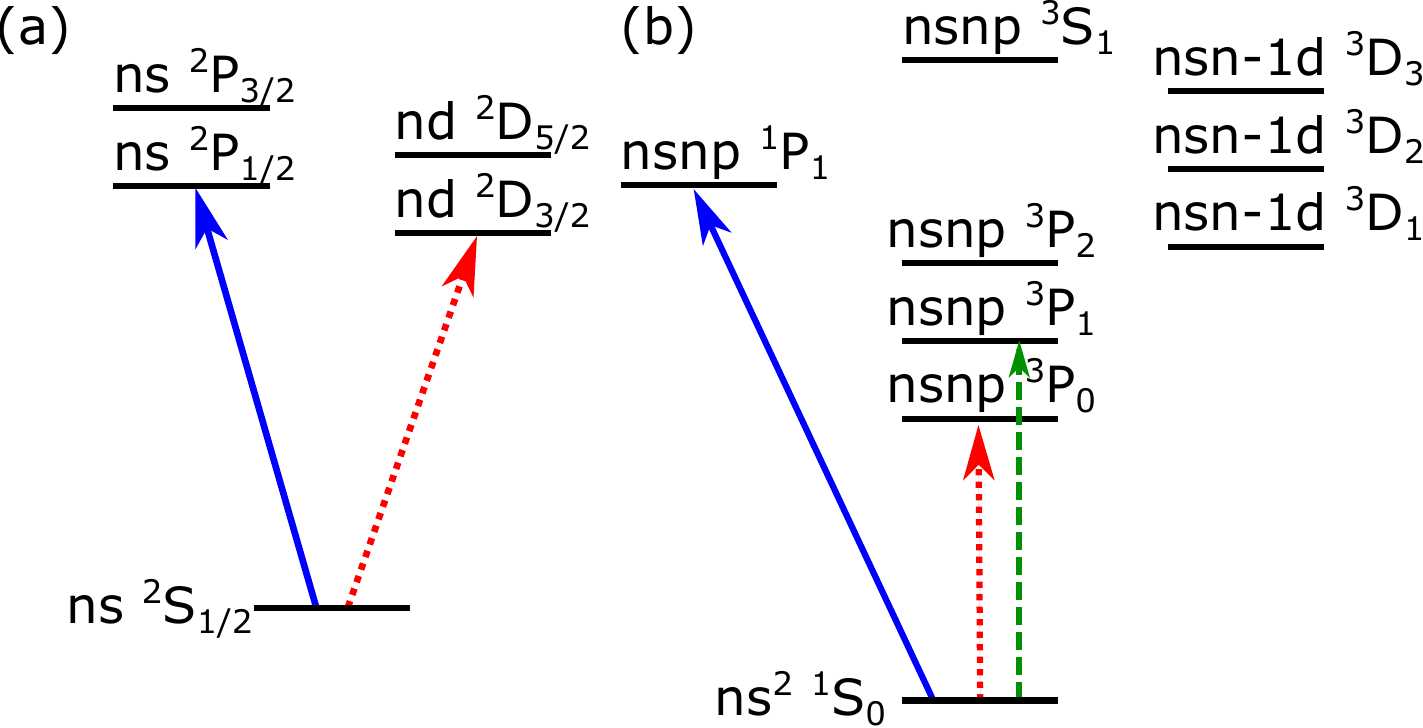}
\caption{Typical energy structure of atoms used in optical atomic clocks: (a) alkali-atom-like structure and (b) alkaline-earth-atom-like structure. The blue solid lines show the transitions used for laser cooling. The green dashed line in (b) is the intercombination transition that is allowed by L-S coupling. The red dotted lines are the clock transitions.}
	\label{FigClockEnergyLevel}
\end{figure}

Alkali-atom-like species have a single electron outside a closed shell and have an energy structure shown in Fig. \ref{FigClockEnergyLevel}(a). A key advantage of these species is the strong transition from the $^2S_{1/2}$ ground state to a $^2P$ excited state suitable for laser cooling. The alkali atoms themselves do not have a narrow-linewidth transition suitable for clock transition, where a clock transition is defined as a transition selected for clock operation. However, some singly-ionized ions of alkaline-earth-atom-like atoms have such narrow-linewidth transitions. In these ions, narrow-linewidth transition used as the clock transition is typically from the $^2S_{1/2}$ ground state to a $^2D$ excited state, corresponding to an electric quadrupole (E2) transition. Examples of ion clocks using this transition are Ca$^+$ \cite{PhysRevApplied.17.034041}, Sr$^+$ \cite{PhysRevLett.131.083002}, Ba$^+$ \cite{PhysRevLett.124.193001}, Ra$^+$ \cite{PhysRevLett.128.033202}, and Hg$^+$ \cite{Science.319.1808}. The linewidth of the E2 transition is typically $\lesssim 2\pi \times 1$ Hz, as listed in Table \ref{TableClockTransition}. An exception is Yb$^+$, which also has an electric octupole (E3) transition to an $^2F$ state, in addition to the two E2 transitions to $^2D$ states. Because the E3 transition is a higher-order effect than an E2 transition, it has an exceptionally narrow linewidth of $2\pi \times 3.20(16)$ nHz \cite{PhysRevLett.127.213001}. 

\subsubsection{Alkaline-earth-atom-like species}\label{SecAEAtoms}
Alkaline-earth-atom-like atoms have two valence electrons outside of a closed shell. In the ground state, all electronic spins pair up, resulting in zero electronic angular momentum. Compared to alkali atoms, these species offer a wider range of electronic states, as shown in Fig. \ref{FigClockEnergyLevel}(b), some of which are advantageous for atomic clocks. The ground state is a singlet state that can be excited to the $^1P_1$ singlet state via a broad-linewidth E1 transition. This transition is used for the first stage of laser cooling. Among the triplet states, the $^3P_1$ state has a total angular momentum of $J=1$, allowing transitions from the ground state that satisfy the selection rule $\Delta J=0, \pm1$, except that the transition is between a singlet and a triplet state. Such intercombination transitions are weakly allowed due to spin-orbit (L-S) coupling, which results in a significantly narrower linewidth than the transition to the $^1P_1$ state. The $^1S_0 \rightarrow ^3P_1$ transitions in some atoms, such as Sr and Yb have narrow linewidth convenient for cooling atoms down to $\sim10$ $\upmu$K or lower, leading to the use in the second stage of laser cooling. 

The transition from the ground state to the $^3P_0$ state is strictly forbidden because both states have $J=0$. However, if the nucleus has a nonzero magnetic moment, it induces slight mixing between the $^3P_0$ state and the $^3P_1$ state, making the transition probability finite. The resulting linewidth is as narrow as $\sim 2 \pi \times 1$ mHz, as summarized in Table \ref{TableClockTransition}. Also, thanks to the absence of electronic angular momentum, the transition is insensitive to magnetic fields, which is advantageous for stability under ambient fields. Because of these features, the $^1S_0\rightarrow ^3P_0$ transition is widely used as the clock transition. Note that without a nuclear magnetic moment, a strong magnetic field must be applied to induce mixing at a level comparable to the case with a nuclear magnetic moment \cite{PhysRevLett.96.083001}. Experimentally, Rabi frequencies of $O(1)$ Hz are reported for Sr\cite{PhysRevA.81.023402} and Yb\cite{PhysRevA.77.050501} with a magnetic field of a few mT and a clock laser intensity of $O(100)$ mW/cm$^2$.

\begin{table}[!t]
\centering
	\caption{Major optical clock transitions: $\lambda_{\rm trap}$ is the magic wavelength for optical lattice clocks. The transitions listed here have sufficiently small uncertainty to be included in the secondary representation of the second. The units for $\nu_0 $ and $\lambda_{\rm trap}$ are THz and nm, respectively. Note that the clock transition for $^{88}$Sr is strictly forbidden and is induced by applying a strong magnetic field along with the resonant light field. }
	\label{TableClockTransition}	
\begin{tabular}{llrrr} 
\hline
Atom	& Transition	& $\nu_0$ & $\Gamma /2\pi$ & $\lambda_{\rm trap}$ \\
\hline
$^{27}$Al$^+$ 	& $3s^2~^1S_0\rightarrow 3s3p~^3P_0$					& 1121 	& 7.73(53) mHz \cite{PhysRevLett.98.220801}		& 	  \\ 
$^{40}$Ca$^+$ 	& $4s~^2S_{1/2}\rightarrow 3d~^2D_{5/2}$			& 411  	& 0.186 Hz \cite{MeasurementSensors.18.2665}		& 	  \\
$^{88}$Sr$^+$ 	& $5s~^2S_{1/2} \rightarrow 4d~^2D_{5/2}$			& 445  	& 0.4 Hz \cite{JPhysConfSe.723.012018}	&  	  \\
$^{171}$Yb$^+$	& $6s~^2S_{1/2} \rightarrow 5d~^2D_{3/2}$			& 688  	& 2.58(27) Hz \cite{JPhysB.48.065003}	&  	  \\
$^{171}$Yb$^+$	& $6s~^2S_{1/2} \rightarrow 4f^{13}6s^2~^2F_{7/2}$	& 642  	& 3.20(16) nHz \cite{PhysRevLett.127.213001}	& 	  \\
$^{199}$Hg$^+$	& $6s~^2S_{1/2} \rightarrow 5d ~^2D_{5/2}$			& 1065 	& 1.83(8) Hz \cite{PhysRevA.42.5425}		& 	  \\
$^{87}$Sr		& $5s^2~^1S_0 \rightarrow 5s5d~^3P_0$				& 429	& 1.35(3) mHz \cite{PhysRevResearch.3.023152}	& 813 \\
$^{88}$Sr 		& $5s^2~^1S_0 \rightarrow 5s5d~^3P_0$				& 429  	& 	& 813 \\
$^{171}$Yb		& $6s^2~^1S_0 \rightarrow 6s6p~^3P_0$				& 518  	& 7.03(6) mHz \cite{PhysRevLett.113.033003}	& 759 \\
$^{199}$Hg		& $6s^2~^1S_0 \rightarrow 6s6p~^3P_0$		 		& 1129 	& 0.1 Hz \cite{RevModPhys.87.637}		& 363 \\
\hline
\end{tabular}
\end{table}

The $^3P_2$ state also has a sufficiently narrow linewidth to serve as a clock transition, because the transition from the ground state to the $^3P_2$ state is the magnetic quadrupole transition. However, it has not been used as intensively as the $^3P_0$ state for clock operations. Instead, this state is used as a probe for many-body physics in quantum degenerate gases \cite{PhysRevA.96.032711}. Yb has one more metastable state: the $4f^{13}5d6s^2(J=2)$ state. This state is theoretically predicted to be beneficial for various fundamental physics searches \cite{PhysRevA.98.022501,PhysRevLett.120.173001}. Recently the direct transitions to this state from the ground state \cite{PhysRevA.107.L060801,PhysRevLett.130.153402} and the $6s6p~^3P_0$ state \cite{PhysRevX.14.011023} have been observed for the first time. Although the uncertainty in its absolute frequency measurement still needs improvement before it can be used as a clock transition, an initial analysis of its potential application in nuclear and particle physics has already been published \cite{PhysRevA.109.062806}. 

Many atomic species with alkaline-earth-atom-like structure are used in atomic clocks. Among neutral atoms, most alkaline-earth-atom-like atoms have been experimentally investigated: Mg \cite{PhysRevLett.115.240801}, Ca \cite{Metrologia.44.146}, Sr \cite{PhysRevLett.133.023401}, Cd \cite{PhysRevLett.123.113201}, Yb \cite{Nature.564.87}, and Hg \cite{PhysRevLett.108.183004}. In the case of Ba the $^3D$ states lie at lower energies than the $^3P$ states, preventing the $^3P_0$ state from serving as a clock transition. So far, only a theoretical proposal is available for Be \cite{NewJPhys.25.043011}. The $^3P$ states in Ca have exceptionally a narrow linewidth, and in the past, the $^3P_1$ state was investigated as a clock transition. Sr and Yb are the most common atomic species for optical lattice clocks. Recently, the narrow-linewidth transitions in these atoms have also been applied to atom interferometry \cite{QuantumSciTech.6.044003} and quantum computation \cite{PhysRevX.8.041054,PhysRevX.8.041055}. Some ion clocks also have alkali-earth-atom-like electronic structure. Examples are Al$^+$~\cite{PhysRevLett.123.033201}, In$^+$~\cite{PhysRevLett.134.023201}, and Lu$^+$~\cite{SciAdv.9.eadg1971}.

\subsection{Optical lattice clock}
An optical lattice clock is an atomic clock where thousands of neutral atoms are trapped in an optical lattice. The trapping laser is tuned to a special frequency called the magic frequency (often referred to as the magic wavelength when describing the corresponding wavelength) \cite{FreqStdMetrol.323,PhysRevLett.91.173005}. At this frequency, the ac Stark shift caused by the trapping laser is cancelled out between the ground and excited states of the clock transition. Most atoms suitable for optical lattice clocks have alkaline-earth-atom-like structure, as they have a narrow-linewidth transition in neutral atoms. 

The stability of the state-of-the-art optical lattice clock reached $1.6 \times 10^{-18}$ in 2013 \cite{Science.341.1215}. In this report, the stability at 1 s was $3.2\times 10^{-16}$, and by integrating over $\sim 10^4$ s, the stability reached the low $10^{-18}$ range. This high stability is achieved by a combination of low QPN of $1\times10^{-16} $ with $N\sim5000$ and $\tau_{R}=140 $ ms. To further improve the SQL, increasing $\tau_{R}$ is crucial, which requires improving the stability of the clock laser. Efforts to develop more stable clock lasers are ongoing using cryogenic cavities \cite{PhysRevLett.119.243601,Optica.6.240}, achieving a fractional stability of $6.5\times10^{-17}$ from 0.8 s to 80 s, or 16 mHz linewidth for 1542 nm laser. Another cryogenic silicon cavity for 1397 nm laser reached a stability of $4.3(2)\times10^{-17}$ from 4 to 12 s, corresponding to a linewidth of 9.6(3) mHz linewidth \cite{SciBull.70.3337}. 

A recently developed technique for quickly estimating systematic shifts is the differential measurement of multiple atomic ensembles with a common clock laser. The advantage of this method is that the differential response of atoms is immune to laser fluctuation and noise, including the Dick effect \cite{Dick1987,IEEETransUlrtasonFerroelectricsFreqControl.45.887}, enabling precise comparisons even with a less stable laser. This method can be implemented various ways, such as two atomic ensembles in separate vacuum chambers\cite{NatPhoton.11.48}, multiple atomic ensembles in the same vacuum chamber\cite{Nature.602.425}, and a single atomic ensemble analyzed in two subregions using camera imaging \cite{Nature.602.420}. The highest reported stability at 1 s is $1.5\times 10^{-18}$ \cite{PhysRevLett.135.103601}, while the record long-term stability of an atomic ensemble is $7.6\times 10^{-21}$ \cite{Nature.602.420}. Higher stability allows for the estimation of systematic uncertainties in a shorter time, contributing to improving clock accuracy. An optical lattice clock with a fractional uncertainty of $2\times 10^{-18}$ was reported in 2015 \cite{NatCommun.6.6896}, and now the best performing clock has a fractional uncertainty of $8.1\times 10^{-19}$~\cite{PhysRevLett.133.023401}. 

This method of comparing multiple atomic ensembles is used to investigate higher atomic coherence time. Best coherence time achieved is longer than 100 s. Ref. \cite{PhysRevLett.135.103601} made use of low-density atoms trapped in a shallow optical lattice to suppress the decoherence due to the Raman scattering of the excited state by the lattice light and atomic collision. Ref. \cite{2505.06437} utilized hyperfine-resolved readout to enhance the coherence. 

\subsubsection{Major systematic uncertainties}
For an optical atomic clock to be highly accurate, it is essential to accurately estimate systematic shifts. Optical lattice clocks are typically stable enough that statistical uncertainty can be minimized by increasing the overall measurement time. As a result, their overall uncertainties are primarily limited by systematic uncertainties. Historically, improvements in the accuracy of optical lattice clocks have been enabled by a deeper understanding of systematic shifts. The pursuit of better uncertainty estimates have motivated the development of techniques to further mitigate these shifts. A notable  example is the dc Stark shift. When optical lattice clocks were reported for the first time, the dc Stark shift caused by stray electric fields in a standard vacuum chamber were not a major contributor to total systematic uncertainty. However, as the overall uncertainty of optical lattice clocks improved, the impact of ambient electric fields became increasingly significant\cite{IEEETransUFFC.59.411}. To precisely quantify this effect, the dc Stark shift was carefully measured with a chamber with electrodes inside \cite{PhysRevLett.120.183201}. In addition, to prevent charge accumulation on insulator surfaces, the inner surface of the viewports has conductive coating, typically indium tin oxide, which helped suppress stray electric fields.

One of the major sources of uncertainty in optical lattice clock is the black body radiation (BBR) shift. This uncertainty arises from two factors: the amount of BBR experienced by the atoms and the uncertainty in the polarizability of the ground and, particularly, the excited state. To estimate the polarizability precisely, the transition matrix for the $^3P_0\rightarrow ^3D_1$ transition, which has the smallest transition frequency among the E1 transitions from the $^3P_0$ state, was carefully measured for Sr \cite{PhysRevLett.133.023401} and Yb \cite{PhysRevA.86.051404}. To control the amount of BBR, a metallic shield with stabilized temperature is implemented inside the vacuum chamber. The temperature is calibrated to the 1-mK level for Sr\cite{NatCommun.6.6896} and Yb\cite{Nature.564.87}, confirmed by a Pt thermometer located near the atoms. In addition, the inner surfaces of both the vacuum chamber and the shield are coated with carbon nanotube to keep the emissivity of the chamber wall as close to 1 as possible \cite{PhysRevLett.113.260801}. Surrounding atoms with a cold shield inside a vacuum chamber reduces the amount of BBR at the atom location, as dictated by the Stefan-Boltzmann law. Such a cryogenic optical lattice clock for Sr at 95 K achieved a small BBR uncertainty \cite{NatPhoton.9.185}, with thorough uncertainty budget listed. A Yb optical lattice clock with a cryogenic shield is recently reported \cite{PhysRevLett.135.063402}. The temperature is set at 77 K, and achieved fractional uncertainty for the BBR shift is $1.7 \times 10^{-20}$. 

A smaller uncertainty from the BBR shift can also be achieved by using atoms less sensitive to BBR. Hg, Cd, Mg, and Tm are attractive for this reason \cite{ApplPhysLett.120.140502}. Clocks based on Hg\cite{PhysRevLett.108.183004}, Cd\cite{PhysRevLett.123.113201}, and Tm\cite{JETPLett.119.659} have already been realized, while a Mg clock is under development, with the clock transition frequency and the magic wavelength already measured \cite{PhysRevLett.115.240801}. 

Another major source of uncertainty originates from the ac Stark shift caused by the trapping laser (Ref.\cite{PhysRevLett.133.023401} for Sr, Ref.\cite{Nature.564.87} for Yb). When the total uncertainty was on the order of $10^{-16}$, the shift was mitigated by carefully selecting magic wavelength for Sr\cite{PhysRevLett.98.083002} and Yb\cite{PhysRevLett.103.063001}. However, with an orders of magnitude smaller uncertainty, higher order effects must also be taken into account. One such effect arises from vector and tensor shifts, which depends on the polarizations of the trapping lasers\cite{PhysRevLett.106.210801}. The shift proportional to the trap depth $U_0$ is formulated as 
\begin{equation}\label{EqACStark1}
\Delta \nu^{E1} = \left( \Delta \kappa^s + \Delta \kappa^v m_F \xi \hat{\bf k} \cdot \hat{\bf B} + \Delta \kappa^t \beta \right) U_0,
\end{equation}
where $\beta=(3|\hat{\bf \varepsilon} \cdot \hat{\bf B}|^2-1)[3m_f^2-F(F+1)]$, with $\hat{\bf \varepsilon}$, $\hat{\bf k}$, and $\xi$ being the lattice polarization vector, lattice propagation vector, and lattice polarization ellipticity, respectively, and $\hat{\bf B}$ being the direction of the quantization axis. $\Delta \kappa^s$, $\Delta \kappa^v$ , and $\Delta \kappa^t$ are the differential scalar, vector, and tensor shift coefficients, respectively. For a one-dimensional lattice, properly selecting the relative orientation between the polarization and the magnetic field can cancel some of these shifts. However, in a three-dimensional lattice, such cancellation is not possible in all three dimensions. This limitation is one of the reasons why state-of-the-art optical lattice clocks use a one-dimensional optical lattice. 

The ac Stark shift $\Delta \nu^{LS}$ including nonlinear terms for a one-dimensional optical lattice is characterized as follows \cite{PhysRevLett.121.263202}:
\begin{align}\label{EqACStark2}
\frac{\Delta \nu^{LS} (u,\delta_L,n_z)}{\nu_0} =& \left( \frac{ \partial \tilde{\alpha}_{E1}}{\partial \nu_L} \delta_L -\tilde{\alpha}_{M1E2} \right) \left( n_z + 1/2 \right) u^{1/2}  \nonumber \\
&-\left[ \frac{ \partial \tilde{\alpha}_{E1}}{\partial \nu_L} \delta_L + \frac{3}{2} \tilde{\beta} \left( n_z^2+n_z+ \frac{1}{2} \right) \right]u \nonumber \\
&+ 2 \tilde{\beta} \left( n_z+\frac{1}{2} \right) u^{3/2} -\tilde{\beta}u^2,
\end{align}
where $u=U_0/E_R$ with $E_R$ being recoil energy for lattice light, $\tilde{\alpha}_{E1}$ is the differential E1 polarizability between the ground and excited states of the clock transition, $\nu_L$ is the lattice laser frequency, $\delta_L$ is the detuning of the lattice laser frequency from the E1 magic frequency and $\tilde{\alpha}_{M1E2}$ and $\tilde{\beta}$ are the differential multipolarizability and hyperpolarizability, respectively. This formulation accounts for the effect of the longitudinal temperature of atoms in the lattice via $n_z$ and incorporates the shifts in Eq. \ref{EqACStark1}. The radial temperature $T_r$ can be included by modifying $u^j$ as $(1+jk_BT_r/uE_R)^{-1}u^j$, where $j$ is the exponent on $u$ for each term in Eq. \ref{EqACStark2}. Using this formalism, the concept of operational magic intensity was itroduced \cite{PhysRevLett.121.263202}. However, despite these characterizations, the lattice ac Stark shift remains one of the major sources of uncertainty in the best optical lattice clocks. A careful analysis on the order of $10^{-19}$ is reported for Sr \cite{PhysRevLett.130.113203}, leading to a fractional uncertainty of $8.1\times 10^{-19}$ \cite{PhysRevLett.133.023401}. Similar analysis is also going on for Yb \cite{PhysRevLett.134.033201}.
An alternative approach to suppress the uncertainty from the ac Stark shift is to use atoms with small higher-order effects. Particularly, hyperpolarizability for Hg, Cd, and Zn is calculated to be smaller than that for Sr and Yb \cite{PhysRevA.93.043420}. These clocks can be potentially useful to suppress uncertainties related to the lattice ac Stark shift, in addition to the BBR shift. 

The ac Stark effect also shifts the transition frequency due to the clock laser itself. Because the clock laser must be applied to drive the clock transition, minimizing this shift requires long irradiation times with low and uniform intensity. For such operations, inducing Rabi flopping is more effective than using a Ramsey sequence, which requires a separate interrogation time in addition to $\pi/2$ pulses. As a result, in state-of-the-art optical lattice clocks, a single $\pi/2$ pulse is commonly used to measure the frequency offset of the clock laser from the atomic resonance. 

The density shift, which arises from atom-atom interactions, is another source of uncertainty. Since the shift is proportional to the density of atoms in the lattice, reducing the density helps mitigate its effects. Another way to control the density shift is by trapping atoms in a three-dimensional optical lattice and removing excess atoms by lowering the Fermi temperature of a degenerate Fermi gas\cite{Science.358.90}. However, both approaches have drawbacks. Reducing the density of atoms increases the QPN due to the small number of atoms, while the three-dimensional optical lattice induces a large uncertainty in the ac Stark shift due to the tensor shift. Additionally, the edge effect \cite{PhysRevLett.127.013401} needs to be carefully managed. Recent studies utilize low-density fermionic gas in large vertical one-dimensional optical lattices. First, a large optical lattice, e.g., a beam waist size of 260 $\upmu$m \cite{Nature.602.420}, is used to reduce the density of atoms. To achieve such a lattice, trap laser intensity is enhanced with an intermediate finesse cavity (${\cal F}\sim 200$\cite{PhysRevLett.119.253001}, $1300$\cite{Nature.602.420}). Second, fermionic isotopes naturally suppress the collisional shift by s-wave collisions. For Sr, the remaining shift due to p-wave collisions is canceled by inter-layer s-wave collisions, which are induced by the delocalized atoms over a few layers due to the low trap depth ($U_0<10E_{R}$) \cite{Nature.602.420}. This approach has enabled the clock to reach $<10^{-18}$ fractional uncertainty \cite{PhysRevLett.133.023401}. For Yb, delocalization of the atomic cloud to reduce the density in this large one-dimensional optical lattice is performed. \cite{PhysRevLett.132.133201} Other systematic uncertainties, such as the background gas collision shift, the second order Zeeman shift, and the servo error are well-controlled and significantly suppressed compared to the major uncertainties discussed here. 

\subsubsection{Varieties of optical lattice clocks}
As the optical lattice clock technology advances, systems designed for various applications are also developed. These systems extend beyond the conventional setup, where a single atomic ensemble is loaded into a single location in an optical lattice. 

Transportable optical lattice clocks have been developed for applications such as geodesy testing and clock comparisons at remote locations. One example is a laboratory built on a trailer \cite{PhysRevLett.118.073601}. More compact versions have been used to test the validity of general relativity at an observation tower \cite{NatPhoton.14.411}, and efforts to make it even smaller are ongoing. Clock comparisons across continents using such transportable clocks were recently reported \cite{OptLett.50.646,2410.22973}. 

Clock comparisons are also performed using fiber links. Metrology laboratories around the world are linked via optical fiber networks, with the longest network located in Europe \cite{NatCommun.7.12443,PhysRevApplied.18.054009}. These links enable precise clock comparisons between remote national metrology institutes\cite{PhysRevApplied.18.054009}, and facilitate comparisons between different types of clocks with unprecedented accuracy \cite{Nature.591.564}. They are also used to measure the height difference based on time delay \cite{NatPhoton.10.662}. 

For differential comparisons between atomic ensembles, multiple ensembles can be trapped in a single one-dimensional optical lattice. One approach involves using a moving optical lattice\cite{Nature.602.425}, allowing atoms to be trapped at up to six spatially distinct locations. These relative measurements have been used to test general relativity \cite{NatCommun.14.4886} and enhance clock stability \cite{PhysRevX.14.011006}. In a fixed optical lattice, two ensembles can be trapped by shifting the position of the MOT \cite{QuantumSciTechnol.9.045023}. 

Continuous interrogation helps suppress the Dick effect \cite{Dick1987,IEEETransUlrtasonFerroelectricsFreqControl.45.887}, where pulsed interrogation of the atomic system introduces instability at frequencies corresponding to the period of the experiment cycle. In addition, continuous interrogation enables  feedback on the phase, causing the Allan deviation to scale $\sim1/T$ rather than $\sim1/\sqrt{T}$. This allows for a faster estimation of systematic shifts. A continuous generation of an ultracold atomic beam has been reported as a step toward developing a continuous optical lattice clock \cite{PhysRevApplied.21.034006}.

The rapidly advancing technology of optical tweezer arrays is also applied to Sr optical atomic clocks \cite{Science.366.93,PhysRevLett.122.173201}. A stability study using $^{88}$Sr atoms has been reported\cite{Nature.588.408}, though characterization of the uncertainties has not been documented. The ability to control individual atoms is advantageous for generating entangled states, which can further enhance stability, as discussed in Sec. \ref{SecEnhancedClockByEntanglement}.

\subsubsection{Enhancement of stability by entanglements}\label{SecEnhancedClockByEntanglement}
The atomic clock system is one of the few systems where entanglement has been demonstrated to enhance their performance. The first proof-of-principle experiments were performed with microwave clocks using $^{87}$Rb \cite{PhysRevLett.104.250801}, where atoms trapped in an optical cavity are entangled through a strong interaction with photons circulating in the optical cavity. Later, larger stability enhancement was achieved with a system where atoms were uniformly coupled to a light field inside an optical cavity \cite{Nature.529.505}. These studies utilized spin-squeezed states as the source of entanglement. The first spin-squeezed optical atomic clock was demonstrated with Yb\cite{Nature.588.414}. Spin squeezing was performed on the spin-1/2 system in the ground state of $^{171}$Yb \cite{PhysRevLett.122.223203} and then transferred to the excited state of the clock transition to improve clock performance. The study reported a 4.4 dB improvement below the SQL in the stability of the atomic system with a fractional stability of $\sim 10^{-14}$, limited by the frequency fluctuations of the clock laser.  An improvement in the clock stability was demonstrated at the stability of $\sim10^{-17}$ \cite{NatPhys.20.208} with a Sr optical lattice clock. With improved stability of the system, longer interrogation time and improved metrological gain is achieved in the same system. This resulted in the spin squeezed optical lattice clock beyond the SQL at the fractional stability of $1.1 \times 10^{-18}$ \cite{PhysRevLett.135.193202}. Note that all spin squeezing described here was performed in a cavity quantum electrodynamics (QED) system, where atoms were located in a high-finesse cavity that enhances their coupling to light. 

Enhanced stability of optical atomic clocks using Greenberger-Horne-Zeilinger (GHZ) states has been demonstrated with a tweezer array clock.\cite{Nature.634.315}  The GHZ state is generated through a strong atom-atom interaction via a Rydberg state. These strongly entangled atoms exhibit high phase sensitivity, leading to improved clock performance when the Ramsey sequence is used. However, because the GHZ and other entangled states are more sensitive to decoherence, the atomic coherence time decreases, resulting in shorter interrogation time compared to state-of-the-art optical lattice clocks. This negates the advantage of the reduced QPN. The overall stability of the clock does not necessarily improve with entangled states. Nevertheless, the shorter interrogation time decreases the duration of a single clock cycle, increasing the bandwidth of clock measurements. This increase can be beneficial when the interrogation time is constrained, such as in fountain clocks, or in searches for ultralight dark matter with relatively large masses (see Section \ref{SecUltralightDM}).

\subsection{Ion clocks}
\subsubsection{State-of-the-Art Systems}
In a conventional ion clock, a single ion is trapped in a Paul trap, which uses an oscillating quadrupole electric field to confine charged particles\cite{PaulTrap,ZPhys.156.1,RevModPhys.62.531}. Since Paul traps have been in use for a longer time than magic-wavelength optical lattices, ion clocks were developed as the frequency standard before optical lattice clocks, as shown in Fig. \ref{FigClockStability}. Although the one-second stability of the ion clock is not as high as that of optical lattice clocks, their ability to operate over long periods allows their accuracy to be comparable to that of optical lattice clocks.

The ionic species with the smallest fractional uncertainties have traditionally been Al$^+$ \cite{PhysRevLett.123.033201} and Yb$^+$ \cite{PhysRevLett.116.063001}. Lu$^+$ \cite{SciAdv.9.eadg1971}, In$^+$ \cite{PhysRevLett.134.023201}, Ca$^+$\cite{2506.17423}, and Sr$^+$\cite{PhysRevAppl.24.044082} recently reached the fractional uncertainties of a similar level. Al$^+$, In$^+$, and Lu$^+$ have alkaline-earth-atom-like electronic structures, while Yb$^+$, Ca$^+$, and Sr$^+$ have alkali-atom-like structures. Yb$^+$ clocks make use of the E3 transition, where an electron in the $4f$ orbital is excited. Sr$^+$ and Ca$^+$ utilize E2 transitions. Other atoms with alkaline-atom-like structure such as Ba$^+$ \cite{PhysRevLett.124.193001} and Ra$^+$ \cite{PhysRevLett.128.033202} have also been investigated. Some of these clocks are summarized in Table \ref{TableClockTransition}. For certain ions, the broad-linewidth transitions suitable for laser cooling have wavelengths too short for readily available laser systems. In such cases, laser cooling is achieved through sympathetic cooling by trapping another ion in the same Paul trap. The preparation and the detection of the quantum state are also carried out using this second ion through the Coulomb coupling of their motional states, which is known as quantum logic spectroscopy \cite{Science.309.749}. For example, in an Al$^+$ clock, the Al$^+$ ion is sympathetically cooled by a Mg$^+$ ion, and quantum logic spectroscopy is performed via the Mg$^+$ ion \cite{PhysRevLett.123.033201}.

\subsubsection{Uncertainties}
State-of-the-art ion clocks have fractional uncertainties around $1\times 10^{-18}$. An Yb$^+$ clock first reached $3\times 10^{-18}$ in 2016\cite{PhysRevLett.116.063001}, and an Al$^+$ clock reached $9.8\times 10^{-19}$ in 2019. In 2025, an In$^+$\cite{PhysRevLett.134.023201} with $2.5\times 10^{-18}$ is reported. Also, three ion clocks achieved fractional uncertainties below $10^{-18}$: $5.5\times 10^{-19}$ for Al$^+$\cite{PhysRevLett.135.033201}, $7.9\times 10^{-19}$ for Sr$^+$\cite{PhysRevAppl.24.044082}, and $4.6\times 10^{-19}$ for Ca$^+$\cite{2506.17423}. Accuracy of these clocks is limited by several factors. One of the major sources of systematic uncertainty is the micromotion-related Doppler shift, which arises from the ion's micromotion in a Paul trap. In some clocks, this uncertainty is the dominant contributor to the overall uncertainties \cite{PhysRevLett.123.033201,PhysRevLett.116.063001,PhysRevLett.131.083002}. Suppressing this effect can be achieved by optimizing the Paul trap's configuration. The high performance of the best Al$^+$ ion clock was made possible by designing a new electrode configuration that minimizes micromotions \cite{PhysRevLett.123.033201,PhysRevLett.135.033201}, offering a significant improvement over previous-generation setups. 

The BBR shift is another dominant source of uncertainty \cite{PhysRevLett.109.203002, PhysRevLett.131.083002, PhysRevApplied.17.034041, PhysRevLett.123.033201, PhysRevLett.116.063001, PhysRevLett.134.023201} and has been extensively studied. For Lu$^+$\cite{NatCommun.9.1650} and Ca$^+$\cite{JPhysB.50.015002}, dedicated studies  evaluating the BBR shift have been published. A cryogenic clock is realized for a Ca$^+$ ion clock \cite{PhysRevApplied.17.034041}. 

With respect to the ac Stark shift, the probe-induced ac Stark shift can also be a significant source of uncertainty\cite{PhysRevLett.116.063001}. This is particularly relevant for ion clocks that use E2 or E3 transitions as the clock transition, as these require much higher laser intensity compared to the $^1S_0\rightarrow ^3P_0$ transitions, which are weakly mixed with an E1 transition by the nuclear magnetic moment. To mitigate the ac Stark shift induced by the clock laser, the hyper-Ramsey sequence, where pulse lengths are adjusted to cancel the shift, is advantageous\cite{PhysRevA.82.011804}. This sequence is particularly beneficial for the spectroscopy requiring high intensity, such as transitions without any mixing with an E1 transition and two-photon transitions. The hyper-Ramsey sequence was experimentally investigated in an Yb$^+$ ion clock using the E3 transition \cite{PhysRevLett.109.213002}. Even with its application in Ref. \cite{PhysRevLett.116.063001}  the probe-induced ac Stark shift remains a major source of the uncertainty. 

Depending on the atomic species and system architecture, other effects can also contribute significantly to uncertainty. The quadratic Zeeman shift is sometimes a major factor\cite{PhysRevLett.123.033201,PhysRevLett.134.023201}. Al$^+$ has a coefficient for the second-order Zeeman shift more than 10 times larger than that of In$^+$ \cite{PhysRevA.100.013409}. In the case of In$^+$ \cite{PhysRevLett.134.023201}, the uncertainty arises from the relatively large uncertainty in the $g$-factor difference. In addition, the In$^+$ clock also identifies thermal time dilation as its major source of uncertainty \cite{PhysRevLett.134.023201}, which can be mitigated by using a narrow-linewidth transition for laser cooling. Other factors, such as background gas collision and servo error, which are common sources of uncertainty in the optical lattice clocks, also contribute to the overall uncertainty. 

The main disadvantage of ion clocks is their large QPN due to the small number of ions in the trap, unlike optical lattice clocks which contain a much larger number of neutral atoms. The ultimate solution to reducing QPN is to increase the number of trapped ions. Multi-ion clocks have been investigated both theoretically and experimentally. Theoretically proposal exist for both Paul traps \cite{ApplPhysB.107.891} and Penning traps \cite{PhysRevA.92.032108}. In Paul traps, a path towards uncertainty below $10^{-18}$ is demonstrated, and the $\sqrt{N}$ scaling of QPN is reported \cite{PhysRevLett.134.023201}. A key source of uncertainty in multi-ion clocks is the quadrupole shift, which arises from electric field gradients due to trap electrodes and Coulomb interaction between ions. As a result, each ion experiences a different quadrupole electric field, leading to inhomogeneous broadening of the transition. This effect is particularly problematic for ions with large quadrupole moments, such as Ca$^+$, Sr$^+$, and Ba$^+$ \cite{PhysRevA.78.022514}. However, Al$^+$ and In$^+$ are much less affected due to their small quadrupole moments \cite{PhysRevA.95.043405}. Suppression of this quadrupole shift has been demonstrated using dynamic decoupling \cite{PhysRevLett.122.223204}. Also, quantum logic spectroscopy has been successfully applied to multiple ions\cite{PhysRevLett.129.193603}. 

With an advent of multi-ion clocks, enhancing the stability of an ion clock by entanglement is attempted \cite{2506.11810}. Because it is easier to achieve maximally entangled states with few ions than many atoms (See Fig. 2 of Ref. \cite{RevModPhys.90.035005}; entanglement at the Heisenberg limit is achieved by trapped ions), and because the leading factor limiting the QPN of ion clocks is the number of ions, enhancement of the phase sensitivity by the entanglement is effective to improve the stability of ion clocks. In Ref. \cite{2506.11810}, a two-ion Ca$^+$ clock is operated with entanglement, and the highest one-second stability is achieved for Ca$^+$ clocks to date. 

When comparing an ion clock with an optical lattice clock, carefully designed sequences can enhance stability. By coherently linking the zero-dead-time interrogation of Yb optical lattice clocks to an Al$^+$ ion clock, the stability of the frequency comparison was improved by an order of magnitude \cite{NatPhys.19.25}. This enhancement was achieved by extending the interrogation time of the Al$^+$ ion clock, thereby improving the SQL. The longer interrogation time was made possible by the long coherence time of the clock laser, which was maintained through frequent phase corrections without deadtime by the interrogation of the Yb optical lattice clocks.

\section{Variation of fundamental constants}\label{SecAlphaVariation}
\subsection{Theoretical basis}
The constancy of fundamental constants is one of the fundamental assumptions in modern physics. If a fundamental constant were to vary over time or space, it would violate the local position invariance, which states that the result of any local non-gravitational experiment is independent of the time and location of the experiment. Local position invariance is part of Einstein's equivalence principle, along with the weak equivalence principle and local Lorentz invariance. The weak equivalence principle asserts the equivalence of gravitational and inertial mass, while local Lorentz invariance states that the outcome of any local non-gravitational experiment is independent of the velocity of the free-falling reference frame where it is performed. Some review papers \cite{ProgPartNuclPhys.112.103772, LivingRevRelativity.9.3} discuss Einstein's equivalence principle in detail.

To observe any variation of fundamental constants, they must be dimensionless, because a reference is always required for such measurements. For example, the time variation of the proton mass $m_p$ or the electron mass $m_e$ cannot be measured independently, but changes in the proton-to-electron mass ratio $\mu=m_p/m_e$ can be tested. Other important dimensionless fundamental constants include coupling constants for fundamental interactions, such as the fine structure constant $\alpha$.

Historically, the possibility of the variation of fundamental constants was first proposed by Dirac\cite{Nature.139.323}. Nowadays, several mechanisms in BSM physics can induce such variations in the coupling constants of fundamental interactions. In theories with extra dimensions, such as Kaluza-Klein models \cite{PhysRevD.21.2167}, superstring theories \cite{PhysRevLett.57.1978}, and brane-world models \cite{JHEP.1999.010}, the fundamental constants remain uniform across all dimensions but appear effectively constant in the observable $(3+1)$-dimensional world. In addition, a field introduced to explain the inflation theory can couple to fundamental constants. This field is typically a scalar field, due to its simplicity and its ability to acquire a vacuum expectation value without violating local Lorentz invariance. \cite{RepProgPhys.80.126902}

Searches for variations of fundamental constants have a long history. Early investigations were performed through astronomical observations \cite{PhysRevLett.19.1294} and studies of the natural nuclear reactor at Oklo\cite{Nature.264.340}. Since than, extensive astronomical searches for variations of fundamental constants have been long ongoing\cite{RepProgPhys.80.126902,LivingRevRelativ.28.6}. Particularly important were observations based on atomic absorption lines that suggested that $\alpha$ may have had a different value in the region of universe with a large redshift $z$. A study in 1999 \cite{PhysRevLett.82.884} claimed $\Delta \alpha/\alpha=-1.1(4)\times 10^{-5}$ over $0.5<z<1.6$. A subsequent report in 2001 \cite{PhysRevLett.87.091301} showed an increased significance with $\Delta \alpha/\alpha=-0.72(18)\times 10^{-5}$ over $0.5<z<3.5$, where $\Delta \alpha$ is the variation of $\alpha$, with no identified systematic effects that could account for the results. Further analyses incorporating additional data have supported spatial variation of $\alpha$ with $\sim4\sigma$ significance (see Ref.\cite{SciAdv.6.eaay9672} and references therein).  However, recent observations at large $z$ themselves are consistent with no variation \cite{SciAdv.6.eaay9672,AstronAstrophys.658.A123,ResAstronAstrophys.24.125012}. If this spatial variation extends to the scale of the solar system, a variation of $\Delta \alpha/\alpha\sim  10^{-19}$ yr$^{-1}$ is expected due to the motion of the solar system in the universe \cite{EurophysLett.97.20006}. Comparisons using atomic clocks provide a valuable cross-check against astronomical searches, as they allow for better control of environment of atoms than astronomical searches.

The search for time variations of fundamental constants on Earth is performed by comparing the transition frequencies of two different clock transitions over a long period. Because the dependence of atomic transitions on fundamental constants varies between transitions, the ratio of their transition frequencies has a nonzero dependence on fundamental constants. Consider $\alpha$ as an example. The binding energies of energy levels in hydrogen with a principal quantum number $n_p$ scale as $\sim \alpha^2$. The fine structure, which arises from L-S coupling, is proportional to $\alpha^4$. As a result, the frequency ratio of these two energy scales is proportional to $\sim \alpha^2$. For precise frequency measurements, narrow-linewidth transitions used in atomic clocks are more suitable than broad-linewidth transitions in hydrogen. In multi-elecron atoms, the $\alpha$ dependence of a transition is estimated by theoretical calculations of the electronic structure and is characterized by a constant $K$ in the following equation:
\begin{equation}\label{EqK}
\frac{\Delta \nu}{\nu_0} = \frac{2q}{\hbar \omega_0}\frac{\Delta \alpha}{\alpha} = K\frac{\Delta \alpha}{\alpha},
\end{equation}
where $q$ is the sensitivity parameter to variations of $\alpha$. 

Table \ref{TableK} summarizes $K$ for major transitions used in atomic clocks, along with some transitions proposed for future measurements that have particularly large K. For neutral atoms and singly-ionized ions, $K$ relative to the ground state is at most $O(1)$. HCIs (see Sec. \ref{SecHCI}) are necessary to obtain $K>10$. The nuclear clock transition in $^{229}$Th is predicted to be highly sensitive to the variation of $\alpha$ \cite{NatCommun.16.9147}. (see Sec. \ref{SecNuclearClock})

\begin{table}[!t]
\centering
	\caption{Sensitivity to the variation of the fine structure constant for major transitions used in atomic clocks. At the end of the table, transitions in HCI with particularly large $K$ and a nuclear clock transition are listed for reference. "Ground state hyperfine" and "RF" for Rb and Cs refers to transitions between hyperfine states in the ground state, with energy differences in the radio frequency range.}
	\label{TableK}
\begin{tabular}{llrrr} 
\hline
Atom	& Transition	& $\lambda $ (nm)	& $K$ & Ref.\\
\hline
Rb 		& ground state hyperfine							& RF & $0.34$ 	& \cite{PhysRevC.73.055501} \\
Cs 		& ground state hyperfine							& RF & $0.83$ 	& \cite{PhysRevC.73.055501} \\
Al$^+$ 	& $3s^2~^1S_0\rightarrow 3s3p~^3P_0$					& 267 & $0.008$	& \cite{PhysRevA.70.014102}  \\
Sr 		& $5s^2~^1S_0 \rightarrow 5s5d~^3P_0$				& 698 & $0.06$ 	& \cite{PhysRevA.70.014102} \\
Sr$^+$ 	& $5s~^2S_{1/2} \rightarrow 4d~^2D_{5/2}$			& 674 & $0.43$ 	& \cite{PhysRevA.59.230} \\
Yb		& $6s^2~^1S_0 \rightarrow 6s6p~^3P_0$				& 578 & $0.31$ 	& \cite{PhysRevA.70.014102} \\
Yb		& $6s^2~^1S_0 \rightarrow 6s6p~^3P_2$				& 507 & $0.56$ 	& \cite{PhysRevA.98.022501} \\
Yb		& $6s^2~^1S_0 \rightarrow 4f^{13}5d6s^2(J=2)$ 		& 431 & $-3.82$	& \cite{PhysRevA.98.022501}  \\
Yb		& $6s6p~^3P_0 \rightarrow 4f^{13}5d6s^2(J=2)$ 		& 1695 & $-15$ 	& \cite{PhysRevLett.120.173001} \\
Yb		& $6s6p~^3P_2 \rightarrow 4f^{13}5d6s^2(J=2)$ 		& 2875 & $-27$		& \cite{PhysRevA.107.053111}
 \\
Yb$^+$	& $6s~^2S_{1/2} \rightarrow 5d~^2D_{5/2}$			& 411 & $1.03$ 	& \cite{PhysRevA.59.230,CanJPhys.87.25} \\
Yb$^+$	& $6s~^2S_{1/2} \rightarrow 5d~^2D_{3/2}$			& 436 & $1.00$ 	& \cite{PhysRevA.59.230,CanJPhys.87.25} \\
Yb$^+$	& $6s~^2S_{1/2} \rightarrow 4f^{13}6s^2~^2F_{7/2}$	& 467 & $-5.95$	& \cite{PhysRevA.77.012515}  \\
Hg$^+$	& $6s~^2S_{1/2} \rightarrow 5d ~^2D_{5/2}$			& 282 & $-2.94$	& \cite{PhysRevA.77.012515}  \\

Pr$^{10+}$	& $5s^25p_{1/2} \rightarrow 5s^24f_{5/2}$		& 2650(49)	& $40$	& \cite{PhysRevLett.113.030801} \\
Pr$^{10+}$	& $5s^25p_{1/2} \rightarrow 5s^24f_{7/2}$		& 1409(14) 	& $22$	& \cite{PhysRevLett.113.030801} \\

Sm$^{14+}$	& $4f^2~^3H_4 \rightarrow 5s4f~^3F_2$		& 2614(470) & $-66$		& \cite{PhysRevLett.113.030801} \\

Hf$^{12+}$	& $4f^{12}(J=6) \rightarrow 4f^{11}5p(J=7)$		& 1777 & $-69.4$		& \cite{2511.00440} \\

Ir$^{17+}$	& $4f^{13}5s~^3F_{4} \rightarrow 4f^{14}~^1S_{0}$		& 1978 & $145$		& \cite{PhysRevLett.106.210802} \\

Cf$^{15+}$	& $6p^25f~^2F_{5/2} \rightarrow 6p^25f~^4I_{9/2}$		& 812 & $57$		& \cite{PhysRevA.92.060502} \\
Cf$^{16+}$	& $5f6p (J=3) \rightarrow 6p^2(J=0)$		& 1899 & $-140$		& \cite{PhysRevLett.109.070802} \\
Cf$^{16+}$	& $5f6p (J=3) \rightarrow 5f^2(J=4)$		& 1030 & $85$		& \cite{PhysRevLett.109.070802} \\
Es$^{16+}$	& $6p5f^2~^4I_{9/2} \rightarrow 6p^25f~^2F_{5/2}$		& 1430 & $-53$		& \cite{PhysRevA.92.060502} \\

$^{229}$Th	& nuclear clock transition		& 148 & $5900(2300)$		& \cite{NatCommun.16.9147} \\

\hline
\end{tabular}
\end{table}

\subsection{Status of experimental searches in laboratories}
\begin{figure}[!tb]
\centering
	\includegraphics[bb=0 0 567 378,width=1\linewidth]{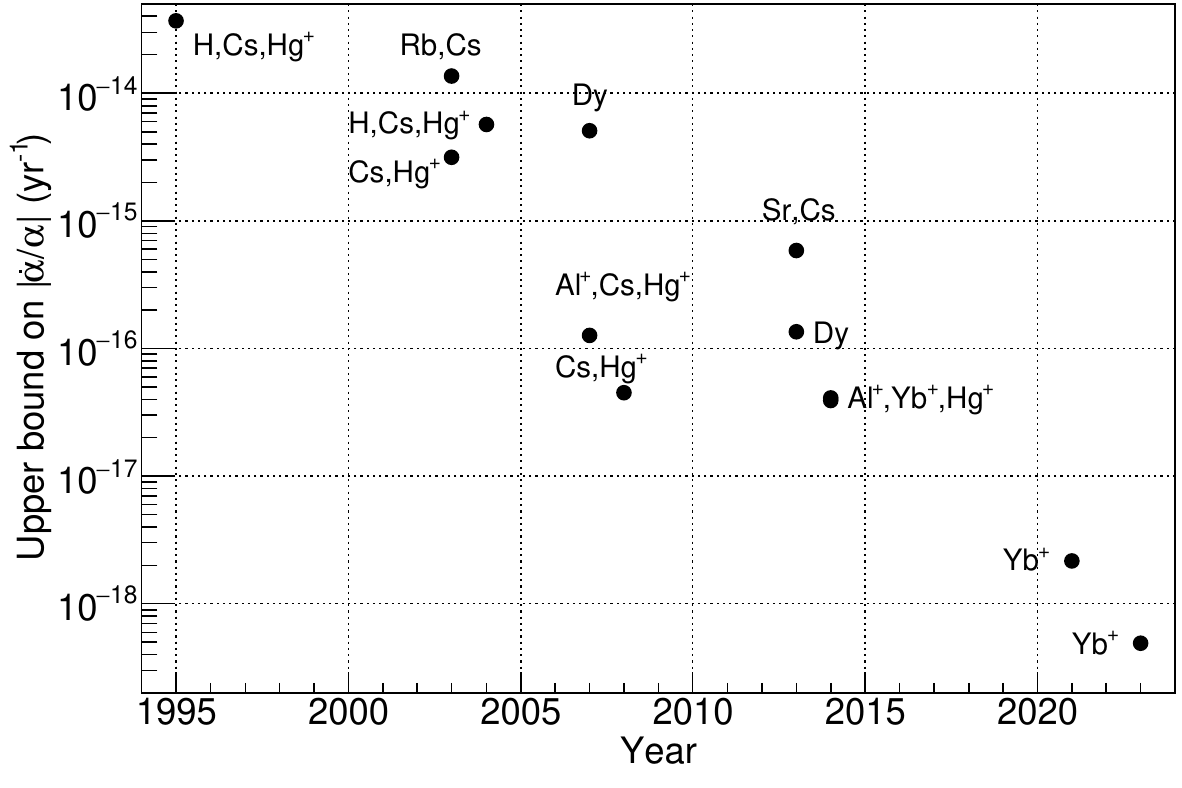}
	\caption{Historical development of the constraint on the time variation of the fine structure constant set by atomic spectroscopy: atomic species next to the data points indicate the species used to obtain the corresponding data points. 
Data are cited from
H,Cs,Hg$^+$ (1995) \cite{PhysRevLett.74.3511},
Rb,Cs \cite{PhysRevLett.90.150801},
Cs,Hg$^+$ \cite{PhysRevLett.90.150802},
H,Cs,Hg$^+$ (2004) \cite{PhysRevLett.92.230802},
Dy (2007) \cite{PhysRevLett.98.040801},
Cs,Hg$^+$ \cite{PhysRevLett.98.070801},
Al$^+$,Cs,Hg$^+$ \cite{Science.319.1808},
Sr,Cs \cite{NatCommun.4.2109},
Dy (2013) \cite{PhysRevLett.111.060801},
Al$^+$,Yb$^+$,Hg$^+$ \cite{PhysRevLett.113.210801,PhysRevLett.113.210802},
Yb$^+$ (2021) \cite{PhysRevLett.126.011102}, and 
Yb$^+$ (2023) \cite{PhysRevLett.130.253001}. Except for Ref.\cite{PhysRevLett.74.3511} these original papers report best fit for the linear drift $\dot{\alpha}/\alpha$ with $1\sigma$ uncertainty ($\sigma$: standard deviation). These drift rates are consistent with zero within $1.96\sigma$ width. $|\dot{\alpha}/\alpha|$ in the plot is 95\% confidence level (C.L.) upper bound calculated as $1.96\sigma$. For Ref. \cite{PhysRevLett.74.3511}, the number reported in the original paper is plotted. 
	}
	\label{FigAlphaHistory}
\end{figure}

The historical advancement in the constraint on the time variation of $\alpha$ by clock comparisons is summarized in Fig. \ref{FigAlphaHistory}. The constraint on the time variation of $\alpha$ is described as the upper bound on $|\dot{\alpha}/\alpha|$. Before the data points shown in Fig. \ref{FigAlphaHistory}, an early report in 1976 set a constraint of $|\dot{\alpha}/\alpha|<4 \times 10^{-12}$ yr$^{-1}$ \cite{Turneaure1976}. Another report combined a clock comparison and astrophysical observations and set the constraint of $|\dot{\alpha}/\alpha|<2.7 \times 10^{-13}$ yr$^{-1}$ in 1993 \cite{PhysRevLett.71.2364}.  Since 1995, the constraint has rapidly improved with the development of atomic clocks. At an early stage, the Hg$^+$ clock played a significant role \cite{PhysRevLett.74.3511,PhysRevLett.90.150802,PhysRevLett.92.230802,PhysRevLett.98.070801,Science.319.1808}, and recent best constraints have been obtained using Yb$^+$ ion clocks \cite{PhysRevLett.113.210801,PhysRevLett.113.210802,PhysRevLett.126.011102,PhysRevLett.130.253001}, both of which have large $K$ (see Table \ref{TableK}). Because the constraints are estimated for a linear drift, frequency ratio measurements over a long period enhance the sensitivity. The recent reports using Yb$^+$ ion clocks utilize the frequency ratio data spanning more than 10 years\cite{PhysRevLett.130.253001}. The best constraint reported so far, ${\dot \alpha}/\alpha=1.8(2.5)\times10^{-19}$ /yr \cite{PhysRevLett.130.253001}, is on the same order of magnitude as the spatial variation reported in astronomical observations, when converted according to the motion of the Earth \cite{EurophysLett.97.20006}.

To increase the sensitivity to the time variation of $\alpha$ in a fixed amount of time, either the stability of an atomic clock or $K$ has to be enhanced. Notably, the $4f^{13}5d6s^2(J=2)$ state in neutral Yb can increase sensitivity \cite{PhysRevA.98.022501,PhysRevLett.120.173001,PhysRevA.107.053111}. Although $K$ for the transition from the ground state to this state is not the largest in Table \ref{TableK}, optical lattice clocks have higher short-term stability than ion clocks, which makes this state beneficial for enhancing sensitivity, especially for short-term variations. Note that the large $K$'s in the $6s6p~^3P_0\rightarrow 4f^{13}5d6s^2(J=2)$ and $6s6p~^3P_2\rightarrow 4f^{13}5d6s^2(J=2)$ transitions are mainly due to their small transition energy, and the excited state is the same as the $6s^2~^1S_0\rightarrow 4f^{13}5d6s^2(J=2)$ transition. In this case, the overall sensitivity for the variation of $\alpha$ needs to be carefully estimated for unstable clocks. Ref. \cite{PhysRevA.98.022501} argues that the enhancement in $K$ is cancelled by the clock instability, if the instability is mainly due to noise on the $4f^{13}5d6s^2(J=2)$. To take advantage of transitions with large $K$, transitions in highly charged ions and $^{229}$Th are promising. 

Time variation of $\mu=m_p/m_e$ is also widely investigated. Similar to $\alpha$, the dependence on $\mu$ varies between transitions and is calculated theoretically. Refs. \cite{PhysRevLett.113.210801,PhysRevLett.113.210802} performed a combined analysis of the constraints on ${\dot \alpha}/\alpha$ and ${\dot \mu}/\mu$, resulting in ${\dot \mu}/\mu=0.2(1.1)\times 10^{-16}$ yr$^{-1}$ and ${\dot \mu}/\mu=-0.5(1.6)\times 10^{-16}$ yr$^{-1}$, respectively, both consistent with zero.  More stringent constraints of ${\dot \mu}/\mu=5.3(6.5)\times 10^{-17}$ yr$^{-1}$ and ${\dot \mu}/\mu=-0.8(3.6)\times 10^{-17}$ yr$^{-1}$
 have been reported from frequency comparisons between Yb, Sr, and Cs clocks \cite{Optica.4.448} and between the E3 transitions in Yb$^+$ ion and a Cs clock \cite{PhysRevLett.126.011102}, respectively. In addition, molecules are used to investigate time variation of $\mu$ because their internal degrees of freedom exhibit different dependence on $\mu$. The energy differences for purely electronic, vibrational, and rotational transitions $E_{\rm el}$, $E_{\rm vib}$, and $E_{\rm rot}$ scale with $\mu$ as follows \cite{NewJPhys.11.055048}:
\begin{align}
E_{\rm el}  &\sim 1			 \nonumber \\
E_{\rm vib} &\sim \mu^{-1/2}  	\\
E_{\rm rot} &\sim \mu^{-1}.	 \nonumber
\end{align}
Laboratory-based studies employed SF$_6$ molecules \cite{PhysRevLett.100.150801} and ultracold KRb molecules \cite{NatCommun.10.3771}. The most stringent constraint based on molecular measurements is ${\dot \mu}/\mu=0.30(1.0)\times10^{-14}$ /yr \cite{NatCommun.10.3771}, which is less precise than the best constraint from the Yb$^+$ ion clock. However, these reports provide a valuable cross-check for atomic clock comparisons.

\section{Spectroscopy of highly charged ions}\label{SecHCI}
\subsection{Properties of highly charged ions}
A major source of systematic uncertainty in optical atomic clocks arises from the ac Stark shift, which is caused by the deformation of the electronic cloud around the nucleus due to external electric fields. This effect can be mitigated by reducing the deformation, which can be achieved by holding electrons more tightly within the atom. Such strong confinement of electrons can be achieved in a highly charged ion (HCI) \cite{RevModPhys.90.045005}.  Qualitatively, by stripping away loosely bound outer-shell electrons, only those with strong binding remain, leading to reduced susceptibility to external electric fields. Optical atomic clocks based on HCIs have been proposed as a next-generation clock capable of surpassing the accuracy limit of both optical lattice clocks and ion clocks \cite{PhysRevA.86.022517}. 

The qualitative behavior of energy levels in HCIs can be understood through the $Z$ dependence of various energy scales, where $Z$ is the atomic number, as summarized in Table \ref{Zscaling}. For hydrogen-like ions, the $Z$ scaling of physical quantities can be derived from the Schr\"odinger equation and other exact solutions. This calculation shows that the energy difference for the $1S-2P$ transition increases proportional to $Z^2$, while the ac Stark shift decreases with $Z^{-4}$ scaling. In HCIs, the $Z$ dependence arises from three different charges, $Z$, $Z_a$, and $Z_{\rm ion}$, where $Z_a$ is the effective charge that maintains the energy level scaling of an external electron in the HCI at $\sim 1/n_p^2$ and $Z_{\rm ion}$ is the charge of the ion. The scaling summarized in Table \ref{Zscaling} is consistent with hydrogen-like ions when $ Z \sim Z_a \sim Z_{\rm ion}$. 

A key trade-off of HCI clocks compared to optical lattice clocks is the number of ions that can be interrogated simultaneously. HCI clocks typically trap a single HCI, resulting in lower clock stability due to QPN. A HCI is more susceptible to deionization from collisions with residual gas than a singly charged ion, which potentially can limit the duration of long-term operation. 

\begin{table}
\centering
	\caption{$Z$ dependence of energy scales in hydrogen-like ions and HCI: the scaling for HCI is cited from Ref. \cite{PhysRevA.86.022517}. More detailed scaling of the Lamb shift in hydrogen-like atoms is provided in Ref. \cite{AtDataNuclDataTables.33.405}.}
	\label{Zscaling}
\begin{tabular}{ccc} 
\hline
					& hydrogen like ion 	& highly charged ion \\
\hline
binding energy		& $\sim Z^{2}$ 			& $\sim Z_a^{-4}$ \\
fine structure		& $\sim Z^{4}$ 			& $\sim Z^2Z_a^3/(Z_{\rm ion}+1)$ \\
hyperfine structure & $\sim Z^{3}$ 			& $\sim ZZ_a^3/(Z_{\rm ion}+1)$ \\
Lamb shift			& $\sim Z^{4}$ 			&  \\
ac Stark shift		& $\sim Z^{-4}$ 		& $\sim Z_a^{-4}$ \\
BBR shift			& $\sim Z^{-4}$ 		& $\sim Z_a^{-4}$ \\
electric quadrupole shift	& $\sim Z^{-2}$	& $\sim Z_a^{-2}$ \\
\hline
\end{tabular}
\end{table}

Beyond their role in next-generation atomic clocks, HCIs are also useful for various fundamental physics searches. One example is testing QED in strong fields. The intense electric field experienced by electrons in innermost orbitals is suitable for probing QED in strong-field regime. Recent reports include spectroscopy of helium-like U$^{90+}$ \cite{Nature.625.673,PhysRevLett.134.153001} and beryllium-like Pb$^{78+}$ \cite{PhysRevLett.135.113001}. In addition, HCIs exhibit higher sensitivity to the variation of $\alpha$ than neutral atoms. This is because high-velocity electrons in HCIs are affected more by relativistic effects, which scale strongly with $\alpha$. Following a pioneering work \cite{PhysRevLett.105.120801},various ionic species have been proposed as suitable candidates for searching for the variation of $\alpha$. Several transitions with large $K$ are listed in Table \ref{TableK}, with additional transitions provided in Table \ref{TableHCI}. Transitions with longer wavelengths tend to have large $K$, as explained by Eq. \ref{EqK}. Another important application is the search for fifth force between an electron and a neutron (See Sec. \ref{SecIsotopeShift} for further details). Due to their significantly different electronic structures compared to neutral atoms, HCIs offer high sensitivity for detecting new forces through isotope shift measurements \cite{PhysRevA.103.L040801}. 

\begin{table}
\centering
	\caption{Hyperfine structures suitable for laser spectroscopy with $I=1/2$. \cite{PhysRevLett.113.233003}}
	\label{HFclock}
\begin{tabular}{lrr} 
\hline
ion				& wavelength $\lambda$ (nm)	& $\Gamma$ (s$^{-1}$) \\
\hline
$^{171}$Yb$^{69+}$ 	& 2160	& 0.43 \\
$^{195}$Pt$^{77+}$ 	& 1080	& 3.4~~ \\
$^{199}$Hg$^{79+}$ 	& 1150	& 2.8~~ \\
$^{203}$Tl$^{80+}$ 	& 338	& 111.2~~ \\
$^{205}$Tl$^{80+}$ 	& 335	& 114.2~~ \\
$^{207}$Pb$^{81+}$ 	& 886	& 6.2~~ \\
\hline
\end{tabular}
\end{table}

\begin{figure}
\centering
	\includegraphics[bb=0 0 499 376,width=0.6\linewidth]{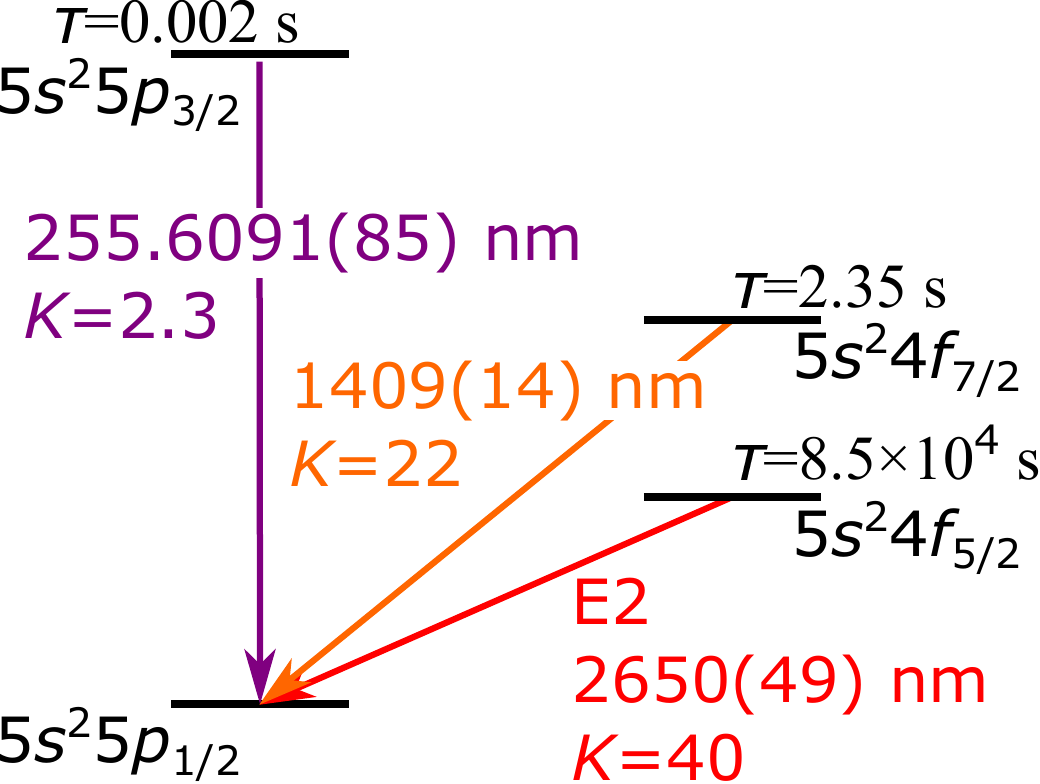}
	\caption{Energy levels for Pr$^{10+}$: E2 shows an electric quadrupole transition. $\tau$ shows the lifetime of the excited states. $K$ is the sensitivity coefficient for the variation of $\alpha$. }
	\label{Praceodymium}
\end{figure}

To perform laser spectroscopy on HCIs, both the atomic species and $Z_{\rm ion}$ must be properly selected. The $Z$ scaling of binding energy suggests that electronic transitions between different $n_p$ states, which typically occur in the visible-light range for neutral atoms, shift into the x-ray region for HCIs. Instead, energy differences between hyperfine or fine structures increase to a range suitable for laser spectroscopy. For hyperfine structures, hydrogen-like ions of heavy atoms exhibit $\sim 1$ eV energy differences. Table. \ref{HFclock} summarizes ions with nuclear spin $I=1/2$ and transition frequencies accessible by lasers. For fine structures, $\sim 1$ eV energy differences are realized at $Z_{\rm ion}\sim10$. An example is the energy-level structure of Pr$^{10+}$, as shown in Fig. \ref{Praceodymium}, where the 255.6 nm transition is between two fine-structure levels. The two $5s^24f$ states are another pair of fine structures. The transition between J=3 and J=2 states in the $5f6p$ state of Cf$^{16+}$, listed in Table \ref{TableHCI}, is also a transition between fine structures. 

\begin{figure}
\centering
	\includegraphics[bb=0 0 642 747,width=0.75\linewidth]{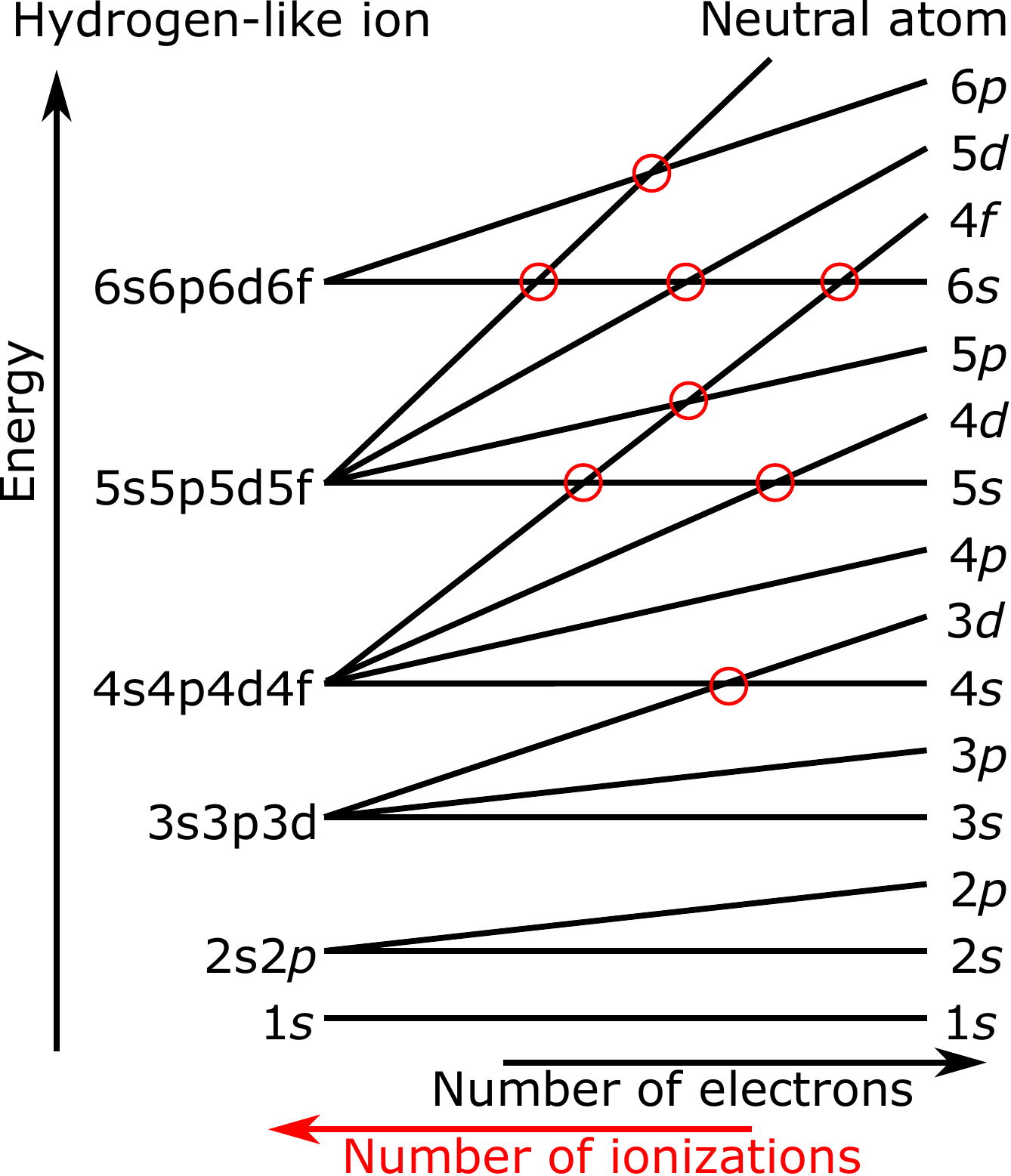}
	\caption{Conceptual illustration of level crossing: the vertical axis shows relative energy. Red circles indicate level crossings. }
	\label{LevelCrossing}
\end{figure}

Naively, energy differences between different orbitals exceed the energy for visible light. However, an exceptional case where visible light can drive electronic transitions in HCIs is known as level crossing \cite{PhysRevA.86.022517}. Figure \ref{LevelCrossing} provides an intuitive illustration of this phenomenon. In a hydrogen atom, different $l$ states for the same $n_p$ level are degenerate, where $l$ is the azimuthal quantum number. However, in multielectron atoms, the $n_p l$ states are ordered from the low energy as $1s,~2s,~2p,~3s,~3p,~4s,~3d,~4p,~5s,~4d,~5p,~6s,~4f,~5d,~6p$. 
Consider, for example, the energy of the $5p$ and $4f$ orbitals. In a hydrogen-like ion, the $5p$ orbital has a higher energy than the $4f$ orbital, whereas in a neutral atom, the $4f$ orbital has higher energy. At a certain $Z_{\rm ion}$ in HCIs, the energy difference between these two states reverses sign. Around this $Z_{\rm ion}$, the energy of the two orbitals is nearly the same, as shown in Fig. \ref{LevelCrossing}, sometimes falling into the visible light range. 

In the case of Pr, level crossing occurs around $Z_{\rm ion}=10$. As shown in Fig. \ref{Praceodymium}, Pr$^{10+}$ exhibits two low-energy transitions between the $5s^25p$ ground state and the $5s^24f$ states, in addition to a transition between fine-structure levels. The two $5s^24f$ states have long lifetimes, making them suitable for clock applications. Transitions from these states to the ground state have high sensitivity to the variation of $\alpha$ (see Table \ref{TableK} for a comparison between neutral atoms and singly-ionized ions).

\begin{table*}[!t]
\centering
	\caption{Narrow-linewidth transitions in HCIs potentially useful for clock operation and the search for variation of $\alpha$: configurations 1 and 2 are the electronic configurations of the two states connected by the transition. J shows the total angular momentum (expressed as a number), or angular momentum in spectroscopic notation. $\lambda$ and $\tau$ are the wavelength and lifetime of the transition, respectively. $K$ is the sensitivity coefficient for the variation of $\alpha$ defined in Eq. \ref{EqK}.}
	\label{TableHCI}
\begin{tabular}{lllllccrr} 
\hline
Atom	& configuration 1 & J		& configuration 2 & J	& $\lambda$ (nm)	& $\tau$ (s)	& $K$ & Ref.\\
\hline

Pr$^{10+}$	& $5p$	& 1/2 	& $ 4f$	& 5/2					& 2650(49)	& $8.5\times 10^4$	& $40$		& \cite{PhysRevA.110.042823} \\
Pr$^{10+}$	& $5p$	& 1/2 	& $ 4f$	& 7/2					& 1409(14) 	& $2.35$	& $22$		& \cite{PhysRevA.110.042823} \\
Nd$^{10+}$	& $4f^2$& $^3H_4$	& $5p4f$ &$^1D_2$			& 2200(430)	& $25$	& $-24$		& \cite{PhysRevLett.113.030801} \\	
Nd$^{10+}$	& $4f^2$& $^3H_4$	& $5p4f$ &$^3F_3$			& 1480(240)	& $3.9$	& $-18$		& \cite{PhysRevLett.113.030801} \\
Pm$^{14+}$	& $5s$ 	& 1/2 	& $4f$	& 7/2					& 966 		& 		& $24$		& \cite{PhysRevLett.105.120801} \\
Sm$^{13+}$	& $5s^24f$	& $^2F_{5/2}$	& $5s4f^2$	&$^4H_{7/2}$	& 494(22) 	& 0.367	& $12$		& \cite{PhysRevLett.113.030801} \\
Sm$^{13+}$	& $5s^24f$	& $^2F_{5/2}$	& $5s4f^2$	&$^4H_{9/2}$	& 444(18) 	& 0.133	& $11$		& \cite{PhysRevLett.113.030801} \\
Sm$^{13+}$	& $5s^24f$	& $^2F_{5/2}$	& $5s4f^2$	&$^4H_{11/2}$	& 386(14) 	& 0.141	& $10$		& \cite{PhysRevLett.113.030801} \\	
Sm$^{14+}$	& $4f^2$	&$^3H_4 $		& $ 5s4f$	&$^3F_2$		& 2614(470)	& 8.515	& $-66$		& \cite{PhysRevLett.113.030801} \\
Sm$^{14+}$	& $4f^2$	&$^3H_4 $		& $ 5s4f$	&$^3F_4$		& 1182(100)	& 0.556	& $-29$		& \cite{PhysRevLett.113.030801} \\
Eu$^{14+}$	& $4f^25s$	& 7/2	& $4f^3$	& 11/2					& 1657(330)	& 		& $47$		& \cite{PhysRevA.90.042513} \\	
Eu$^{14+}$	& $4f^25s$  & 7/2 	& $4f^3$	& 13/2					& 971(130)	& 		& $28$		& \cite{PhysRevA.90.042513} \\
Ho$^{14+}$	& $5s4f^6$	&$^8F_{1/2} $	& $ 4f^55s^2$&$^6H_{5/2}$	& 420 		& 37	& $-16$		& \cite{PhysRevA.91.022119} \\
Hf$^{12+}$	& $5s^24f^{12}$	& 6 	& $5s^24f^{11}5p$ & 7	& 1777 		& 442	& $-69.4$		& \cite{2511.00440} \\
W$^{8+}$	& $4f^{14}5p^4$ &$^3P_{2} $	& $4f^{13}5p^5$ &$^3F_{4} $	& 1646 		& 	& $-27$		& \cite{PhysRevLett.106.210802} \\
W$^{8+}$	& $4f^{14}5p^4$ &$^3P_{2} $	& $4f^{13}5p^5$ &$^3G_{3} $	& 1573 		& 	& $-26$		& \cite{PhysRevLett.106.210802} \\
W$^{8+}$	& $4f^{14}5p^4$ &$^3P_{2} $	& $4f^{13}5p^5$ &$^3G_{5} $	& 899 		& 	& $-15$		& \cite{PhysRevLett.106.210802} \\
Ir$^{16+}$	& $4f^{13}5s^2$ &$^2F_{7/2} $	& $4f^{14}5s$&$^2S_{1/2}$	& 267 	& 		& $22$		& \cite{PhysRevLett.106.210802} \\
Ir$^{17+}$	& $4f^{13}5s$&$^3F_{4}$	& $4f^{12}5s^2$	&$^3H_{6}$	& 283 		& 		& $-22$		& \cite{PhysRevLett.106.210802} \\
Ir$^{17+}$	& $4f^{13}5s$&$^3F_{4}$	& $ 4f^{14}$&$^1S_{0}$		& 1978 		& 		& $145$		& \cite{PhysRevLett.106.210802} \\	
Pt$^{17+}$	& $4f^{13}5s^2$&$^2F_{7/2} $	& $ 4f^{14}5s$&$^2S_{1/2}$	& 402.2	& $4.05\times 10^9$	& $-33$		& \cite{PhysRevA.94.032504} \\
Pu$^{8+}$	& $6p^6$& 0	& $ 6p^55f$&1					& 654		&		& $11$		& \cite{PhysRevA.110.012801} \\
Pu$^{9+}$	& $6p^5$& 3/2	& $ 6p^45f$ & 5/2			& 1172		& 		& $19$		& \cite{PhysRevA.110.012801} \\
Pu$^{9+}$	& $6p^5$& 3/2	& $ 6p^45f$ & 7/2			& 872		& 		& $14$		& \cite{PhysRevA.110.012801} \\
Pu$^{9+}$	& $6p^5$& 3/2	& $ 6p^45f$ & 3/2			& 862		& 		& $14$		& \cite{PhysRevA.110.012801} \\
Pu$^{10+}$	& $6p^4$& 2		& $ 6p^35f$ & 3				& 1989		& 		& $25$		& \cite{PhysRevA.110.012801} \\
Pu$^{10+}$	& $6p^4$& 2		& $ 6p^35f$ & 2				& 866		& 		& $11$		& \cite{PhysRevA.110.012801} \\
Pu$^{10+}$	& $6p^4$& 2		& $ 6p^35f$ & 4				& 817		& 		& $11$		& \cite{PhysRevA.110.012801} \\	
Pu$^{11+}$	& $6p^25f$&5/2	& $ 6p^3$	& 3/2		& 1461		& 		& $-15$		& \cite{PhysRevA.110.012801} \\
Am$^{9+}$	& $6p^55f$& 1 	& $ 6p^45f^2$& 0				& 974		& 		& $12$		& \cite{PhysRevA.110.012801} \\
Am$^{9+}$	& $6p^55f$& 1 	& $ 6p^45f^2$& 4				& 905		& 		& $13$		& \cite{PhysRevA.110.012801} \\
Bk$^{15+}$	& $6p^2$	& 0	& $ 6p5f$ &3						& 483		& 		& $14$		& \cite{PhysRevA.110.012801} \\
Bk$^{15+}$	& $6p^2$	& 0	& $ 6p5f$ &2						& 340		& 		& $13$		& \cite{PhysRevA.110.012801} \\
Cf$^{15+}$	& $6p^25f$&$^2F_{5/2} $	& $ 6p^25f$&$^4I_{9/2}$		& 812		& 6900	& $57$		& \cite{PhysRevA.92.060502} \\
Cf$^{16+}$	& $5f6p$& 3		& $ 6p^2$&0						& 1899		& $3.7\times 10^{16}$	& $-140$		& \cite{PhysRevLett.109.070802} \\
Cf$^{16+}$	& $5f6p$& 3		& $ 5f6p$&2						& 1638		& 0.178	& $35$		& \cite{PhysRevLett.109.070802} \\
Cf$^{17+}$	& $5f$&5/2 		& $ 6p$	&1/2						& 535		& 		& $-48$		& \cite{PhysRevLett.109.070802} \\
Es$^{16+}$	& $6p5f^2$&$^4I_{9/2} $	& $ 6p^25f$&$^2F_{5/2}$		& 1430		& 16000	& $-53$		& \cite{PhysRevA.92.060502} \\
Es$^{17+}$	& $5f^2$&$^3H_{4} $	& $ 6p5f$&$^3F_{2}$				& 1343		& 11000	& $-13$		& \cite{PhysRevA.92.060502} \\

\hline
\end{tabular}
\end{table*}

\subsection{Experimental methods for precision spectroscopy of HCI}
Early HCI spectroscopy was performed by classifying the astronomical observations and detecting radiation in fusion researches, including tokamak experiments \cite{PhysScr.24.663}. These spectroscopic studies primarily focused on the X-ray region and involved mixtures of multiple ionic species. To enable spectroscopy of specific ions, the development of ion sources was essential. The earliest sources of HCIs were based on electron-cyclotron resonance ion sources \cite{ApplPhysLett.16.401}, followed by the invention of electron beam ion sources\cite{NIM.124.157}.  Early optical spectroscopy of HCIs was performed in 1987 for transitions between fine structures \cite{JOSAB.4.144}, and in 1994 for a transition between a hyperfine structure \cite{PhysRevLett.73.2425}. 

Nowadays, HCIs are generated by electron beam ion traps (EBITs) \cite{PhysScr.T22.157}. An EBIT is an electromagnetic ion trap formed by a strong electron beam and electrodes. The negative charge of the electron beam confines the ions radially, while axial confinement is achieved by electrodes. The energy of the electron beam determines the maximum number of electrons that can be stripped from neutral atoms. 

The generated HCIs are extracted from an EBIT and then transferred to a Paul trap. At this stage, the HCIs are hot and must be cooled toward its motional ground state for precision spectroscopy. Due to the absence of broad-linewidth transitions suitable for laser cooling in HCIs, cooling is typically achieved through sympathetic cooling. This method, along with high-precision spectroscopy, has been demonstrated in Ar$^{13+}$ \cite{RevSciInstrum.86.103111}. In this experiment, an Ar$^{13+}$ ion is stopped by a Coulomb crystal of Be$^+$, and subsequently laser-cooled using a broad transition in Be$^+$. The experiment is conducted in a cryogenic trap to suppress the BBR shift \cite{RevSciInstrum.90.073201}. The $^2P_{1/2}\rightarrow^2P_{3/2}$ transition at 441 nm serves as the clock transition, with its transition frequency measured  to a fractional uncertainty of $1.5 \times 10^{-16}$, referenced to the state-of-the-art $^{171}$Yb$^+$ clock. The dominant source of uncertainty in this measurement is the statistical uncertainty, while the total systematic uncertainty for the Ar$^{13+}$ system is $2.2\times 10^{-17}$. The primary contributor to systematic uncertainty arises from micromotions, which can be suppressed in the future work \cite{Nature.611.43}. 

Other ionic species are also investigated, motivated by theoretical proposals. One of the extensively investigated species is Pr ions. Pr$^{9+}$ was first experimentally examined in Ref.\cite{NatCommun.10.5651}. Although $K=6.32(5.28)$ for the $5p4f~^3G_3(^3F_2)$ state is smaller than $K$'s listed in Table \ref{TableHCI}, energies of the long-lived $5p4f~^3G_3$ states have been determined with a $\sim 10^{-6}$ fractional uncertainty. Pr$^{10+}$ also has an early spectroscopy report \cite{PhysRevA.110.042823}. Combined with an improved theoretical calculation of the energies of the metastable excited states \cite{PhysRevLett.113.030801}, the energy of the $5s^25p_{3/2}$ state was determined with a $3\times 10^{-7}$ fractional uncertainty. Transition frequency measurements were also performed for a narrow-linewidth transitions in Ni$^{12+}$ \cite{PhysRevResearch.6.013030,PhysRevLett.135.093002} and many charge states for Xe \cite{PhysRevLett.131.161803}, Ca \cite{PhysRevA.103.L040801}, and Os \cite{2509.06710} ions. An observed narrow-linewidth transition in Ni$^{12+}$ has an expected linewidth of 8 mHz, and the search was performed by a close collaboration between a theoretical prediction of the transition energy and an experimental search for it \cite{PhysRevLett.135.093002}. The investigations of Xe, Ca, and Os ions are motivated by the search for a fifth force through isotope shift measurements (see Sec. \ref{SecIsotopeShift}). In particular,  Ca$^{14+}$ was recently utilized to set a new constraint on the fifth force\cite{PhysRevLett.134.233002}.

\section{Nuclear clock}\label{SecNuclearClock}
\begin{figure}
\centering
	\includegraphics[bb=0 0 587 387,width=0.8\linewidth]{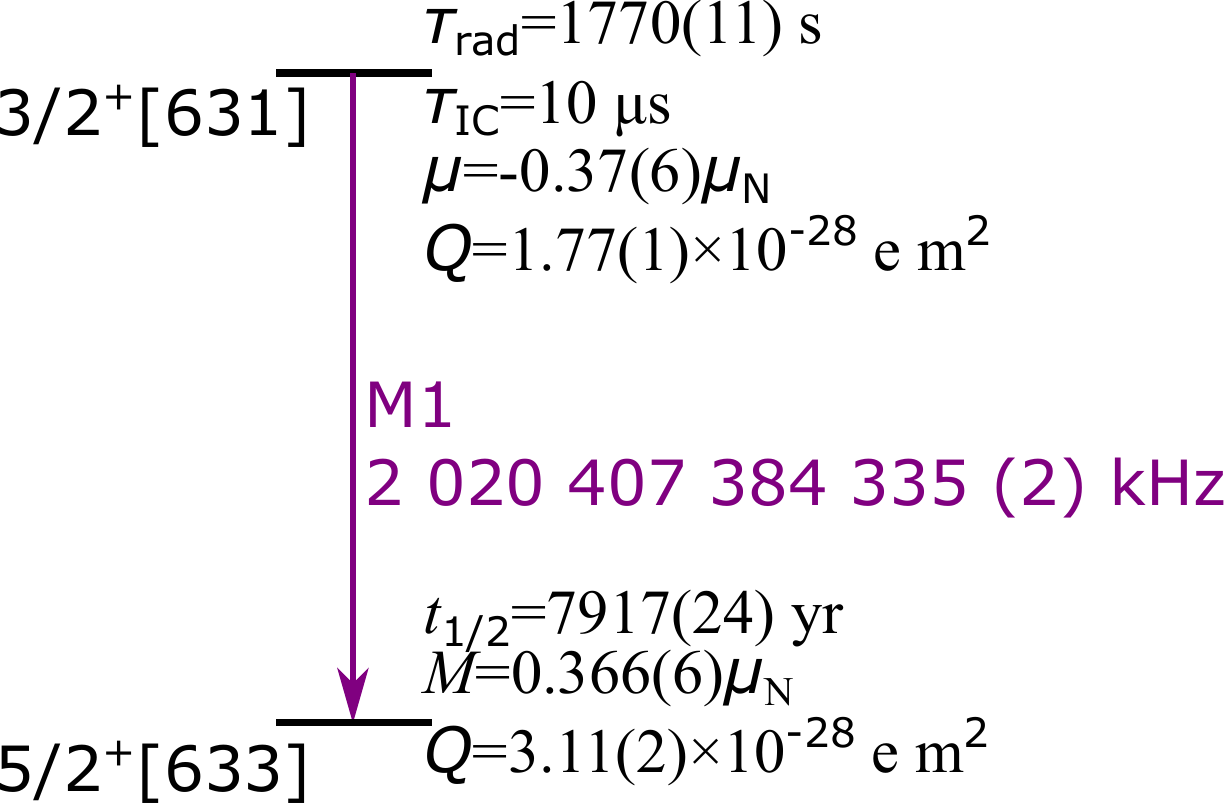}
\caption{Details of the nuclear transition in $^{229}$Th: $\tau_{\rm rad}$: the lifetime for the radiative decay, $\tau_{\rm IC}$: the lifetime for the internal conversion, $M$: the magnetic moment, $\mu_{\rm N}$: the nuclear magneton, $Q$: the quadrupole moment, $t_{1/2}$: the half lifetime. The notation of the angular momentum follows $J^P[N_Dn_{Dz} \Lambda]$, where $J$ is the total angular momentum, $P$ is the parity, $N_D$ and $n_{Dz}$ are the deformed oscillator quantum number and its $z$ component, and $\Lambda$ is the projection of the valence neutron's orbital angular momentum on the nuclear symmetry axis.}
	\label{229Th}
\end{figure}

An object even more rigid against external electric fields than the inner-shell electronic cloud is the nucleus. However, using a typical nuclear transition at $\sim 100$ keV for precision spectroscopy with a laser is extremely challenging. Exceptionally, a low-lying nuclear isomer state in $^{229}$Th is barely accessible by laser excitation from the ground state. The existence of this extremely low-energy nuclear excited state was first proposed by Kroger and Reich \cite{NuclPhysA.259.29}. The idea of using this isomer state as a clock was initially proposed in 2003 \cite{EurophysLett.61.181}. At the time, the energy of this transition was known with only one-digit precision and was estimated to be lower than its currently known value. However, in recent years, substantial advancements have been made. The transition frequency is now determined to be 2 020 407 384 335(2) kHz, corresponding to a wavelength of 148.38 nm and energy of 8.3557 eV, with a fractional uncertainty of $9.9\times10^{-13}$. Additional details of the transition are summarized in Fig. \ref{229Th}. Note that the next lowest-energy nuclear excited state known today is in $^{235m}$U, which is 76.7 eV above the ground state. With this state in $^{229}$Th, a clock can be operated based on a nuclear transition. Because the nucleus is $\sim10^5$ times smaller than the atom, it is nearly unaffected by the ac Stark shift. 


Historical development of the energy estimates for the first excited state of $^{229}$Th is summarized in Table \ref{Table229Th}. Initially, the energy of the excited state was measured by low-energy gamma rays emitted during the decay of $^{233}$U to $^{229}$Th \cite{PhysRevLett.64.271,PhysRevC.49.1845,PhysRevLett.123.222501,PhysRevLett.125.142503}. More recent, precise measurements utilized internal conversion and x-ray excitation of the nucleus to a higher excited state than the one involved in the clock transition \cite{Nature.573.243}. In addition, spectroscopy of photons emitted during the relaxation of the clock state further reduced the uncertainty in the transition energy\cite{Nature.617.706}, improving the precision of the transition frequency to four digits. A breakthrough occurred when laser excitation of the transition and subsequent decay of the excited state was observed in $^{229}$Th$^{4+}$ ions doped in a CaF$_2$ crystal \cite{PhysRevLett.132.182501}. Soon after that a similar observation was reported with Th-doped LiSrAlF$_6$ \cite{PhysRevLett.133.013201}. These studies measured the wavelength with seven digit precision, confirming that the observation was not crystal specific but resulted from $^{229}$Th interacting with the light field. A few months later, spectroscopy using the seventh harmonic generation of a frequency comb determined the transition energy with 2 kHz uncertainty \cite{Nature.633.63}. The observed lifetime of the excited state is 1770(11) s. In the field of optical lattice clocks and ion clocks, well-established methodologies can reduce the uncertainty of resonant frequency measurements from 1 kHz to $<1$ Hz. However, a key difference between these optical atomic clock systems and the $^{229}$Th nuclear clock is the lack of a stable continuous-wave laser operating at the resonant frequency of the clock transition. 

With an observation of the transition and the determination of the transition frequency, the investigation is moved to the further characterization of the transition. While this long lifetime provides excellent coherence for long interrogation time, it also makes quick manipulation challenging. To address this, methods to shorten the lifetime were explored, using laser-induced quenching \cite{PhysRevResearch.7.L022036} and a thin film of the host material \cite{Nature.636.603}. Further reduction of the lifetime of the isomer state is achieved by performing spectroscopy in a material opaque to the light resonant to the nuclear clock transition \cite{2506.03018}. ThO$_2$ has a band gap of 6 eV, smaller than the nuclear transition at 8.4 eV. This allows decays through internal conversion, reducing the lifetime to $\sim10$ $\upmu$s. Further studies have also investigated thermal effects \cite{PhysRevLett.134.113801} and reproducibility in the transition frequency between different batches of doped crystals at the temperature where the first order thermal sensitivity is zero \cite{2507.01180}. A theoretical analysis pointed out that the transition frequency for the nuclear clock transition can be different between host materials, ions with different charge, and bare nuclei \cite{PhysRevLett.135.123001}. 

The nuclear clock can be implemented in two different ways. To prevent the $^{229}$Th nucleus from rapidly decaying via internal conversion, the $^{229}$Th atoms must be ionized into $^{229}$Th$^{3+}$ ions. These ions can be trapped using an ion trap similar to those used in HCI clocks. This approach has an advantage of leveraging well-established spectroscopy techniques from ion clocks, making it easier to estimate systematic uncertainties. However, due to the small number of ions in the system, the QPN is large. Towards this direction, laser spectroscopy of $^{229}$Th$^{3+}$ ions in a Paul trap was demonstrated \cite{Nature.629.66}. Another reports sympathetically cools Th$^{3+}$ ions with $^{88}$Sr$^+$ ions \cite{PhysRevA.109.033116}, and performed spectroscopy on hyperfine structures and isotope shifts in Th$^{3+}$ ions \cite{2504.00974}. An alternative method, unique to nuclear clocks, is doping a crystal with $^{229}$Th$^{3+}$ or $^{229}$Th$^{4+}$ ions. Because nucleus is highly isolated from the surrounding electrons, the effects being in a crystal are minimal compared to those on atoms. This approach allows for a large number of ions, e.g., $4.5\times 10^{15}$ \cite{PhysRevLett.133.013201}, to be incorporated, which helps suppress QPN. However, a recent study showed a strong temperature dependence of the transition frequency \cite{PhysRevLett.134.113801}, attributed to changes in the electric field induced by the thermal expansion of the lattice constant, unless temperature is set at a specific temperature where the first order thermal shift is zero \cite{2507.01180}. 

\begin{table}
\centering
	\caption{Historical development of the frequency estimate for the clock transition in $^{229}$Th.}
	\label{Table229Th}
\begin{tabular}{llr} 
\hline
Year & Energy (eV, otherwise noted)	& Ref. \\
\hline
1976	& <100 						& \cite{NuclPhysA.259.29} \\
1990	& $-1\pm4$					& \cite{PhysRevLett.64.271} \\
1990	& <5						& \cite{PhysRevC.42.R499} \\
1994	& 3.5(1.0)					& \cite{PhysRevC.49.1845} \\
2005	& 5.5(1.0)					& \cite{PhysRevC.71.044303} \\
2007	& 7.6(5)					& \cite{PhysRevLett.98.142501} \\
2019	& 8.28(17)					& \cite{Nature.573.243} \\
2019	& 8.30(92)					& \cite{PhysRevLett.123.222501} \\
2020	& 8.10(17)					& \cite{PhysRevLett.125.142503} \\
2023	& 8.338(24)					& \cite{Nature.617.706} \\
2024	& 8.367(24)					& \cite{NatCommun.15.5536} \\
2024	& 8.355 74(3)				& \cite{PhysRevLett.132.182501} \\
2024	& 8.355 733(10)				& \cite{PhysRevLett.133.013201} \\
2024	& 2 020 407 384 335(2) kHz	& \cite{Nature.633.63} \\

\hline
\end{tabular}
\end{table}

For fundamental physics applications, the nuclear clock is expected to be highly sensitive to variation of fundamental constants. Because nuclear structure depends on the coupling constant of the strong force $\alpha_s$, it should be sensitive to variation of $\alpha_s$ \cite{PhysRevLett.97.092502}. In addition, theoretical predictions suggest that it also exhibits high sensitivity to the variation of $\alpha$, with $K=5700(2300)$ \cite{NatCommun.16.9147}. Intuitively, this large $K$ arises because the small transition energy results from a cancellation between a large repulsive Coulomb energy and an equally large attractive energy from the strong force. The large $K$ also makes the nuclear clock a promising tool for searching for ultralight dark matter (see Sec. \ref{SecUltralightDM}). Another potential research direction involves developing a system that operates without vacuum chamber \cite{ApplPhysRev.126.111101}. Such a system could be advantageous for compact portable optical atomic clock. 

\section{Ultralight dark matter}\label{SecUltralightDM}
\subsection{Theoretical basis}\label{SecULDMTHeory}
Dark matter is massive objects in the universe that do not interact with light. It was first proposed to explain the rotation curves of galaxies \cite{MNRAS.311.441}. Today, multiple observations strongly support its existence, such as gravitational lensing \cite{RepProgPhys.73.086901} and fluctuations in the cosmic microwave background \cite{Planck2018}. Dark matter is estimated to be five times more abundant than ordinary matter. 

Various dark matter candidates have been examined, e.g., massive compact halo objects \cite{AstrophysJLett.824.L31} with astronomical mass scales, weakly interacting massive particle typically searched for in the mass range from 0.1 GeV to 1000 GeV \cite{PhysRevD.110.030001}, and axions and axion-like particles, which are generally investigated in the sub-1 eV mass region \cite{PhysRevD.110.030001}. Among these candidates, particles with masses of 1 eV or smaller have to be bosonic due to their expected number density, the energy density of dark matter, and the Pauli exclusion principle. This class of low-mass dark matter is referred to as ultralight dark matter \cite{PhysRevLett.85.1158,AstronAstrophysRev.29.7}. Searches for ultralight dark matter are conducted using various methods including astronomical observation \cite{PhysRevD.109.103030,PhysLettB.737.30,PhysRep.198.1,JCAP.2019.016,PhysRevLett.128.091102}, microwave technology \cite{PhysRevLett.120.151301,PhysRevLett.126.191802,PhysRevD.92.075012,PhysRevLett.127.081801}, laser-based techniques \cite{EurPhysJC.76.24,PhysRevD.92.092002}, accelerometers \cite{PhysRevD.93.075029,PhysRevLett.120.141101,PhysRevD.105.042007}, and nuclear EDM searches \cite{PhysRevX.7.041034}, among others. Notably, atomic clocks are particularly sensitive to topological dark matter, which consists of spin-0 particles forming topological defects in the universe and coupling to non-derivative terms in the SM Lagrangian\cite{PhysRevD.91.015015}. Note that axions and axion-like particles are also important candidates for ultralight dark matter. However, because these particles are extensively reviewed elsewhere \cite{RevModPhys.97.025005,AnnRevNuclPartSci.65.485,AnnRevNuclPartSci.71.225,PhysRep.870.1}, the discussion here is limited to the spinless particles. 

Searches for ultralight dark matter using atomic clocks are sensitive to mass ranges whose energy scale is the same as or smaller than the single cycle of the clock operation, which is $\sim 1$ s. This corresponds to a dark matter mass of $10^{-15}$ eV/c$^2$ or lower. The lower bound for this mass is determined by the overall duration of clock operation. From a cosmological perspective, dark matter is generally expected to have a mass greater than $10^{-22}$ eV in order to account for all observed phenomena with a single candidate. This is because the de Broglie wavelength of dark matter exceeds the size of dwarf galaxies at this mass. However, if dark matter consists of multiple components, mass ranges below $10^{-22}$ eV may still be relevant for investigation. 

The first proposal for a search for ultralight dark matter using atomic clocks focused on detecting topological dark matter with quadratic coupling to SM particles \cite{NatPhys.10.933}. This concept initially relied on GPS satellites \cite{NatCommun.8.1195} and required collaboration between multiple research groups across different geographic locations for experimental realization \cite{SciAdv.4.eaau4869}.  Later, a new approach was proposed to detect ultralight dilaton dark matter, putting emphasis on the linear coupling to fundamental constants \cite{PhysRevD.91.015015}. This proposal included clock comparisons within a single location. After this proposal, various experimental efforts and theoretical analyses of existing data were conducted to put constraints on the ultralight dark matter, starting with a constraint derived from Dy spectroscopy \cite{PhysRevLett.115.011802}. 

For dark matter that couples to fundamental constants, a scalar field is typically assumed. The interaction between this field and ordinary matter is described by the following Lagrangian density \cite{PhysRevD.91.015015,PhysRevA.94.022111,PhysRevLett.115.201301}: 
\begin{align}\label{EqUltralightDM}
{\cal L}_{\rm int} =& \left( \kappa\phi \right)^n \Biggl[ \frac{d_\alpha}{4e^2}F_{\mu\nu}F^{\mu\nu} -\frac{d_g \beta_g}{2g_3} F^A_{\mu\nu}F^{A\mu\nu} \nonumber \\
& - \sum_{i=e,u,d} (d_{m_i}+\gamma_{m_i} d_g) m_i \overline{\Psi}_i \Psi_i \Biggl].
\end{align}
Here, $\kappa=\sqrt{4\pi G}=1/\sqrt{2} M_P$ with G being the Newtonian constant of gravitation and $M_P$ being Planck mass, $F^{\mu\nu}$ is the electromagnetic field tensor, $F^{A\mu\nu}$ is the intensity tensor for the gluon field,  $g_3$ is the coupling constant for quantum chromodynamics (QCD), $\beta_g$ is the $\beta$ function for the running coupling constant for $g_3$, $m_{e,u,d}$ are the mass of an electron, an up quark and a down quark, respectively, $\gamma_{m_i}$ is the anomalous dimension characterizing  $m_i$, and $\Psi_i$ is the spinor for fermions. $n$ is integer, and $n=1,2$ correspond to linear and quadratic coupling. Some studies parametrize the coupling constant as $\kappa d_i=1/\Lambda_i$, in which case $\Lambda_i$ is associated with the energy scale for new physics \cite{NatPhys.10.933,PhysRevLett.115.201301}. Each term in Eq. \ref{EqUltralightDM} describes a tree-level interaction, where a SM particle emits an ultralight dark matter particle, with the coupling constant for this interaction being $d_i$. Well-known particles that satisfy these properties are the dilaton and moduli. Similar interactions can be assumed for axions when CP symmetry is violated \cite{PhysRevD.109.015005}. 

When $\phi$ couples to ordinary matter, fundamental constants vary according to the coupling constants in the Lagrangian in the following way:
\begin{align}
\alpha(\phi)&=\alpha(1+d_\alpha \phi), \nonumber \\
m_e(\phi)&=m_e(1+d_{m_e} \phi), \nonumber \\
m_q(\phi)&=m_q(1+d_{m_q} \phi), \nonumber \\
\Lambda_{\rm QCD}(\phi)&=\Lambda_{\rm QCD}(1+d_{g} \phi) \nonumber
\end{align}
Here, $q=u,d$ for $m_q$, and $\Lambda_{\rm QCD}$ is the energy scale for QCD, which dominates the mass of a proton. The dark matter field oscillates at a frequency $2\pi f=m_\phi c^2/\hbar$ ($m_{\phi}$: the mass of the ultralight dark matter) with an amplitude $\phi_0$. The corresponding dark matter density is given by $\rho_\phi=c^2 \pi f^2 \phi_0^2/G$. In general, atomic transitions depend on fundamental constants in different ways. These dependences can be numerically calculated for multi-electron systems to evaluate the sensitivity of various transitions to the variation of fundamental constants. 

Searching for ultralight dark matter through this interaction is relatively easy to implement in a laboratory because transition frequencies of atomic transitions and the size of an atom depend differently on fundamental constants \cite{NatAstron.1.0009,PhysRevA.93.063630}. Take $\alpha$ as an example. The wavelength of a narrow-linewidth laser used for precision spectroscopy is stabilized by locking it to a stable reference cavity, where the mirror separation $L$ is determined by the spacer material. Since the laser wavelength is proportional to $L$, the frequency $\nu$ scales as $\nu\propto 1/L \sim 1/a_B \sim \alpha$, where $a_B=\hbar/(m_e c \alpha)$ is the Bohr radius. However, an atomic transition does not generally follow the same $\alpha$-dependence, as discussed in Sec. \ref{SecAlphaVariation}. This difference in $\alpha$-scaling leads to a periodic variation in the atomic resonance relative to the cavity resonance when $\alpha$ couples to the ultralight dark matter field $\phi$. 

Three other fundamental constants can also influence the atomic transition frequency. In general, when the frequencies of two atomic transitions A and B are monitored, their ratio $\nu_A/\nu_B$ is depicted as 
\begin{equation}
\frac{\Delta(\nu_A/\nu_B)}{\nu_A/\nu_B}=k_\alpha \frac{\Delta \alpha}{\alpha}+k_{m_e} \frac{\Delta(m_e/\Lambda_{\rm QCD)}}{m_e/\Lambda_{\rm QCD)}}+k_{m_q}\frac{\Delta(m_q/\Lambda_{\rm QCD)}}{m_q/\Lambda_{\rm QCD)}},
\end{equation}
where $\Delta X$ denotes the change in $X$ and $k_i$ ($i=\alpha,~m_e,~m_q)$ are the sensitivity coefficient for $\alpha,~m_e,~m_q$. Because four fundamental constants can vary while only a single frequency ratio is measured experimentally, setting a constraint on one fundamental constant requires assuming that the coupling constants for the other three are zero. 

\subsection{Status of experimental searches}

\begin{figure*}[!t]
\centering
	\includegraphics[bb=0 0 2268 1701,width=1\linewidth]{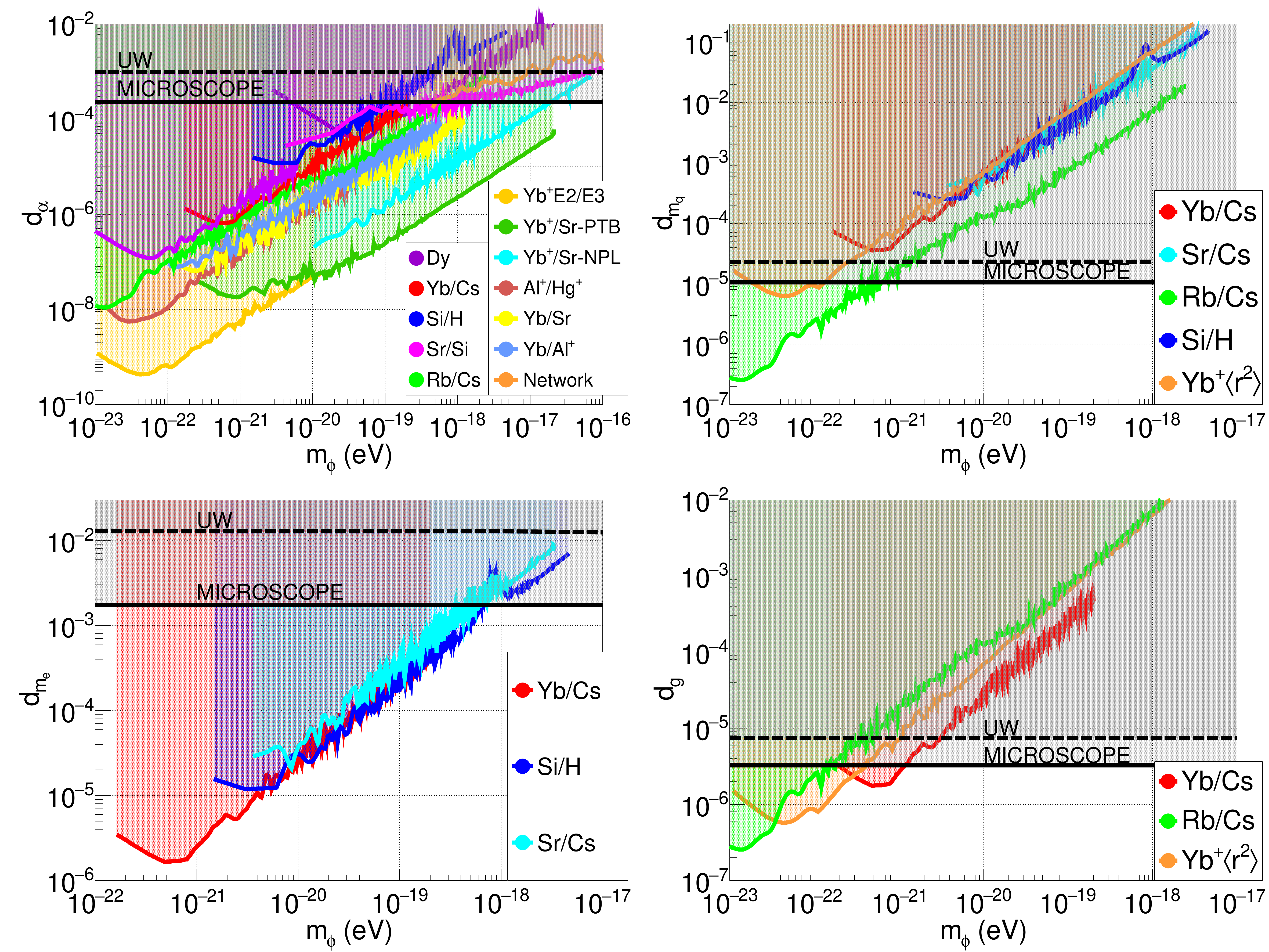}
	\caption{
Constraints on the coupling of ultralight dark matter to (a) the fine structure constant, (b) the electron mass, (c) the quark mass, and (d) the gluon field. Data are cited from 
Dy \cite{PhysRevLett.115.011802},
Yb/Cs \cite{PhysRevLett.129.241301},
Si/H \cite{PhysRevLett.125.201302},
Sr/Si \cite{PhysRevLett.125.201302},
Rb/Cs \cite{PhysRevLett.117.061301},
Yb$^+$ E2/E3 \cite{PhysRevLett.130.253001},
Yb$^+$/Sr-PTB \cite{PhysRevLett.130.253001},
Yb$^+$/Sr-NPL \cite{NewJPhys.25.093012},
Al$^+$/Hg$^+$ \cite{Nature.591.564},
Yb/Sr \cite{Nature.591.564},
Yb/Al$^+$ \cite{Nature.591.564},
Network \cite{SciAdv.4.eaau4869},
Sr/Cs \cite{NewJPhys.25.093012}, 
Yb$^+$ $\langle r^2 \rangle$ \cite{2301.10784}, and 
UW and MICROSCOPE \cite{PhysRevD.98.064051}. In this plot, the final result of MICROSCOPE experiment \cite{PhysRevLett.129.121102} is not included in this plot. Shaded areas are 95\% C.L. excluded region reported by original papers. 
}
	\label{FigDMConstraint}
\end{figure*}

At the end of an experimental analysis, the amplitudes of a sinusoidal oscillation in the frequency ratio are obtained. The frequency $f$ assumed for the sinusoidal fit can be arbitrarily chosen within the range up to the Nyquist frequency. As a result, a single dataset for $\nu_A/\nu_B$ measured over time can provide constraints on dilaton ultralight dark matter across a wide mass range, as shown in Fig.~\ref{FigDMConstraint}. So far, no experimental results detected sinusoidal oscillations with amplitudes significantly larger than their statistical uncertainties. The resulting constraints  exclude coupling constants stronger than a certain threshold for a given mass. In Fig.~\ref{FigDMConstraint}, these excluded regions are represented as shaded areas. In this kind of searches for new particles, it is common practice to plot mass on the horizontal axis and a coupling constant for a specific interaction on the vertical axis. 

Figure~\ref{FigDMConstraint}(a) shows the constraints on $d_\alpha$. Because most electronic transitions exhibit finite sensitivity to the variation of $\alpha$, state-of-the-art atomic clock comparisons have significantly contributed to improving these constraint \cite{Nature.591.564,NewJPhys.25.093012,PhysRevLett.130.253001}. Notably, the result from PTB \cite{PhysRevLett.130.253001} surpasses previous reports by an order of magnitude by combining two pairs of clock comparisons. 
 
Comparisons between optical atomic clocks have no sensitivity to $d_{m_e}$, primarily because all transition frequencies for optical transitions are more or less proportional to $m_e$, resulting in a constant when the frequency ratio is taken. Instead, comparisons between an electronic transition or a reference cavity and a microwave transition between hyperfine levels exhibit a high sensitivity. The comparison between an Si cavity and a hydrogen maser \cite{PhysRevLett.125.201302} reported constraints down to $10^{-21}$ eV. The lower bound on mass is limited by the measurement duration. A long-term comparison between an Yb optical lattice clock and a Cs fountain clock, spanning 298 days, provided constraints for smaller masses \cite{PhysRevLett.129.241301}. 

For $d_g$ and $d_{m_q}$, the contribution of clock comparisons is relatively small compared to satellite \cite{PhysRevLett.119.231101,PhysRevLett.120.141101} and torsion balance \cite{PhysRevLett.100.041101,PhysRevLett.120.141101} experiments, as these macroscopic experiments involve much larger masses. Although Fig. \ref{FigDMConstraint}(c) and \ref{FigDMConstraint}(d) show a slight excess in the performance of the clock comparisons at low masses, the final result of the MICROSCOPE experiment \cite{PhysRevLett.129.121102}, which improved by a factor of 4.6 compared to the initial result shown in the plot, is not yet included. With this updated result, most clock comparison constraints become marginal. Note that the constraint by Yb$^+ \langle r^2 \rangle$ does not follow the formalism discussed in the previous subsection but arises from modulation in nuclear charge radii \cite{2301.10784}.

Constraints on quadratic coupling have been also established through both experimental and theoretical analyses. Early constraints were provided by theoretical studies, where analyses based on Big Bang Nucleosynthesis (BBN) and cosmic microwave background imposed limits several orders of magnitude more stringent than fifth-force searches \cite{PhysRevLett.115.201301}. Experimental constraints from atomic spectroscopy data include measurements from Dy \cite{PhysRevLett.115.201301}, Rb/Cs clock comparison \cite{PhysRevA.94.022111}, Sr clock comparisons between the ground and an observation deck \cite{PhysRevD.102.115016}, and Yb$^+$/Cs clock comparisons \cite{NewJPhys.25.093012}. While constraints from BBN are quite stringent, clock comparisons contribute significantly in the low-mass region for $d_{\alpha}^{(2)}$. 

\subsection{Future proposals}
To increase the search region, three factors are important. A longer measurement time allows for the detection of smaller masses. In principle, ultralight dark matter with arbitrarily small masses can be searched given a sufficiently long measurement period. However, for masses below $10^{-22}$ eV, careful interpretation is required, as discussed in Sec. \ref{SecULDMTHeory}. Stability of clocks is another crucial factor. To investigate smaller coupling constants, the clocks need to be more stable, assuming the same atomic clocks are used. The sensitivity of a transition to the fundamental constant also plays a significant role. Even with the same clock stability, transitions highly sensitive to variations of fundamental constants enhance the overall sensitivity to ultralight dark matter. For example, in the case of $d_\alpha$, transitions with large $K$, as listed in Table \ref{TableK}, are particularly suitable for high-sensitivity measurements. 

Various systems are proposed to be sensitive to searches for ultralight dark matter. One approach is molecular spectroscopy. Microwave spectroscopy of a rovibrational transition in SrOH has a strong sensitivity to $d_{m_e}$ and $d_g$ \cite{PhysRevA.103.043313}. Spectroscopy of Ca$^+$ ion and I$_2^+$ ion is expected to provide a comparable constraint on $d_g$ \cite{PhysRevD.110.015008}. Another proposal involves placing an atomic clock on a satellite orbiting close to the sun \cite{NatAstron.7.113}. The higher density of dark matter bound to the sun enhances sensitivity, particularly improving constraints in the higher mass range, from $10^{-16}$ to $10^{-14}$ eV, beyond existing limits. Other molecular and atomic spectroscopy studies have also placed constraints in the higher mass region \cite{PhysRevLett.129.031302,PhysRevLett.130.251002}. Atom interferometry can contribute within the same mass range as atomic clock comparisons \cite{PhysRevA.110.033313}. The $^{229}$Th nuclear clock is particularly sensitive to ultralight dark matter, with Ref. \cite{PhysRevX.15.021055} discussing its sensitivity $d_g$. In addition, the large $K$ shown in Table \ref{TableK} results in high sensitivity to $d_\alpha$ \cite{,2203.14915}.

\section{Fifth force search with isotope shifts}\label{SecIsotopeShift}
\subsection{Theoretical basis}
Theoretical proposal for searching for new forces using isotope shift measurements emerged around 2018 \cite{PhysRevLett.120.091801,EPJC.77.896}. Around the same time, studies on Higgs coupling to the electron and the up and down quarks \cite{PhysRevD.96.093001}, as well as other new physics searches \cite{PhysRevD.96.015011} were also proposed. Since then, various experimental reports have been published, ranging from simple isotope shift measurements to comprehensive analyses incorporating new isotope shift data, motivated by these theoretical proposals. 

Isotope shifts refer to the frequency difference of an atomic transition between isotopes. This phenomenon was already known in the first half of the 20th century \cite{Nature.104.406,RevModPhys.30.507}. For optical transitions with $\nu_0 \sim500$ THz, isotope shifts $\nu^{AA'}_{i}$ typically range from 10 MHz to 1 GHz. These shifts arise mainly from two sources and can be phenomenologically described as
\begin{equation}\label{EqIsotopeShift}
\nu^{AA'}_{i}= K_{i} \mu^{AA'} + F_{i} \langle r^2 \rangle^{AA'}.
\end{equation}
The first term, referred to as the mass shift, results from differences in nuclear mass. The second term, known as the field shift, is caused by variation in the nuclear charge radius due to differences in the number of neutrons. In Eq. \ref{EqIsotopeShift} and the following discussion, the subscript $i$ labels a transition, and $X^{AA'}=X^A-X^{A'}$, unless otherwise specified. For example, $\nu^{A}_{i}$ is the transition frequency for the transition $i$ of the isotope with mass number $A$. $\mu^{A}=1/m^A$ is the inverse mass of the isotope with mass number $A$, $ \langle r^n \rangle^{A}$ is the $n$th moment of the charge radius for the isotopes with mass number $A$, and $K_i$ and $F_i$ are constants characterizing the mass shift and field shift, respectively. $ \langle r^2 \rangle^{AA'}$ can be eliminated with two isotope shift equations for transitions $i$ and $j$. 
\begin{equation}\label{EqKingPlot}
\frac{\nu_{j}^{AA'}}{\mu^{AA'}}=\frac{F_j}{F_i} \frac{\nu_{i}^{AA'}}{\mu^{AA'}}+K_{j, i} 
\end{equation}
Here, $K_{j,i}=K_{j}-K_{i} F_j/F_i $. The equation shows that $\nu_{j}^{AA'}/\mu^{AA'}$ is a linear function of $\nu_{i}^{AA'}/\mu^{AA'}$. The two-dimensional plot of $\nu_{j}^{AA'}/\mu^{AA'}$ vs $\nu_{i}^{AA'}/\mu^{AA'}$ is called the King plot \cite{JOSA.5.638}. 

Suppose an additional term $X_i^{AA'}$ in Eq.~\ref{EqIsotopeShift}. This term introduces $X_{j,i}^{AA'}/\mu^{AA'}$ on the right-hand side of Eq.~\ref{EqKingPlot}, causing the King plot to become nonlinear. One possible source of this additional term is a hypothetical force between an electron and a neutron, generating the following Yukawa-type potential energy for an electron:
\begin{equation}
V(r)=(-1)^{s+1}\frac{y_ny_e}{4\pi}\frac{e^{-m_\phi c r/\hbar }}{r},
\end{equation}
where $s$ is the spin for the new boson mediating the force, $y_e$ and $y_n$ are the coupling constants of the boson to an electron and a neutron, respectively, and $m_\phi$ is the mass of the new boson. Nonlinearity in a King plot can be attributed to this new force. Conversely, if the plot remains linear, it provides evidence against the existence of such a force. 

Higher-order effects of the nuclear charge radius, which is within the SM, also induce nonlinearity. The following equation is typically assumed in the state-of-the-art analysis:
\begin{align}\label{eqnIS}
\nu_{i}^{AA'}=&F_i \langle r^2\rangle^{AA'}+K_i \mu ^{AA'} + G_i^{(4)}\langle r^4\rangle^{AA'} \nonumber \\
&+ G_i^{(2)}
[\langle r^2 \rangle^2]^{AA'}
+ v_{n e} D_i  N^{AA'}.
\end{align}
The third and fourth terms correspond to the fourth-moment shift and the quadratic field shift, where $[\langle r^2 \rangle^2]^{AA'}=( \langle r^2\rangle^{AA''} )^2 - (\langle r^2\rangle^{A'A''} )^2$ is the second-order perturbative effect due to $ \langle r^2 \rangle^{AA'}$. The equation analogous to Eq.~\ref{EqKingPlot} is 
\begin{align}
\frac{\nu_{j}^{AA'}}{\mu^{AA'}}=&\frac{F_j}{F_i} \frac{\nu_{i}^{AA'}}{\mu^{AA'}}+K_{j, i} +\frac{\langle r^4\rangle^{AA'}}{\mu^{AA'}} G_{j, i}^{(4)}\nonumber \\
&
+\frac{[\langle r^2 \rangle^2]^{AA'}}{\mu^{AA'}} G_{j, i}^{(2)}+\frac{v_{n e} N^{AA'}}{\mu^{AA'}} D_{j, i}. \label{eqnKing}
\end{align}
In this equation, the third, fourth, and fifth terms all contribute to nonlinearity. As a result, properly subtracting background effects from the SM contributions are crucial when searching for the new forces. 

Even with effects that induce nonlinearity, the magnitude of this nonlinearity remained smaller than the uncertainty of the isotope shift measurements when the broad-linewidth transitions were used. This was experimentally tested decades ago\cite{King1984}. Moreover, Eq. \ref{EqIsotopeShift}, which predicts linearity in the King plot, is widely used to determine differences in nuclear root-mean-square (RMS) charge radii \cite{NuclChargeRadii,AtomDataNuclData.99.69}, for heavy atoms by renormalizing all field shifts into a single term by a proportionality coefficient common to all isotopes. One exception is the nonlinearity observed in Sm \cite{JPhysB.14.2769}, which was explained by the field shift \cite{JPhysB.15.997}. With the advancement in precision spectroscopy, as shown in Fig. \ref{FigClockStability}, isotope shift measurements are now sensitive enough to detect nonlinearity beyond the measurement uncertainty caused by the natural linewidth of transitions and other systematic broadening. 

Several conditions are necessary for high-sensitivity measurements. Searching for new forces through King plot nonlinearity requires an element with at least four stable isotopes. The more isotopes and transitions available, the more information can be extracted. Narrow-linewidth transitions are preferred for high-precision measurements. In addition, spinless nuclei are favored to avoid spin-spin interactions that are difficult to estimate precisely. In fact, significant nonlinearity is observed when isotopes with nonzero spin are included in the King plot for Sr \cite{PhysRevResearch.1.033113} and Yb \cite{PhysRevA.109.062806}, with the second-order hyperfine interaction fully explaining the observed nonlinearity in Sr \cite{Phys.Rev.A.112.022804}. For higher sensitivity, transitions involving excited states with large difference in radial wave functions are preferable. For example, a combination of $S\rightarrow D$ and $S\rightarrow F$ transitions is expected to yield higher sensitivity in the King plot compared to a combination of $S\rightarrow ^2D_{5/2}$ and the $S\rightarrow ^2D_{3/2}$ transitions. To suppress unwanted backgrounds arising from SM effects, light nuclei are preferred, because of their small charge radii and reduced relativistic effects. One key source of uncertainty is $\mu^{AA'}$, which can be eliminated by using three narrow-linewidth transitions \cite{PhysRevResearch.2.043444}. Another approach to analyzing isotope shift data is to sort them by isotopes rather than transitions\cite{PhysLettB.838.137682}, which claimed to provide constraints on the new bosons comparable to those obtained through conventional King plot analysis.

This new force coupling to an electron and a neutron is introduced phenomenologically, without necessarily being supported by strong theoretical motivations. However, one benefit of this approach is its sensitivity to the X17 particle \cite{EPJWebConf.232.04005}. The force discussed in this section is spin-independent, while spin-dependent fifth force searches are covered in a recent review paper \cite{RevModPhys.97.025005}. The formalism and recent progress in the King plot analysis are also summarized in another recent review paper\cite{NatRevPhys.7.119}.

\begin{figure}[!tb]
\centering
	\includegraphics[bb=0 0 2268 2265,width=1\linewidth]{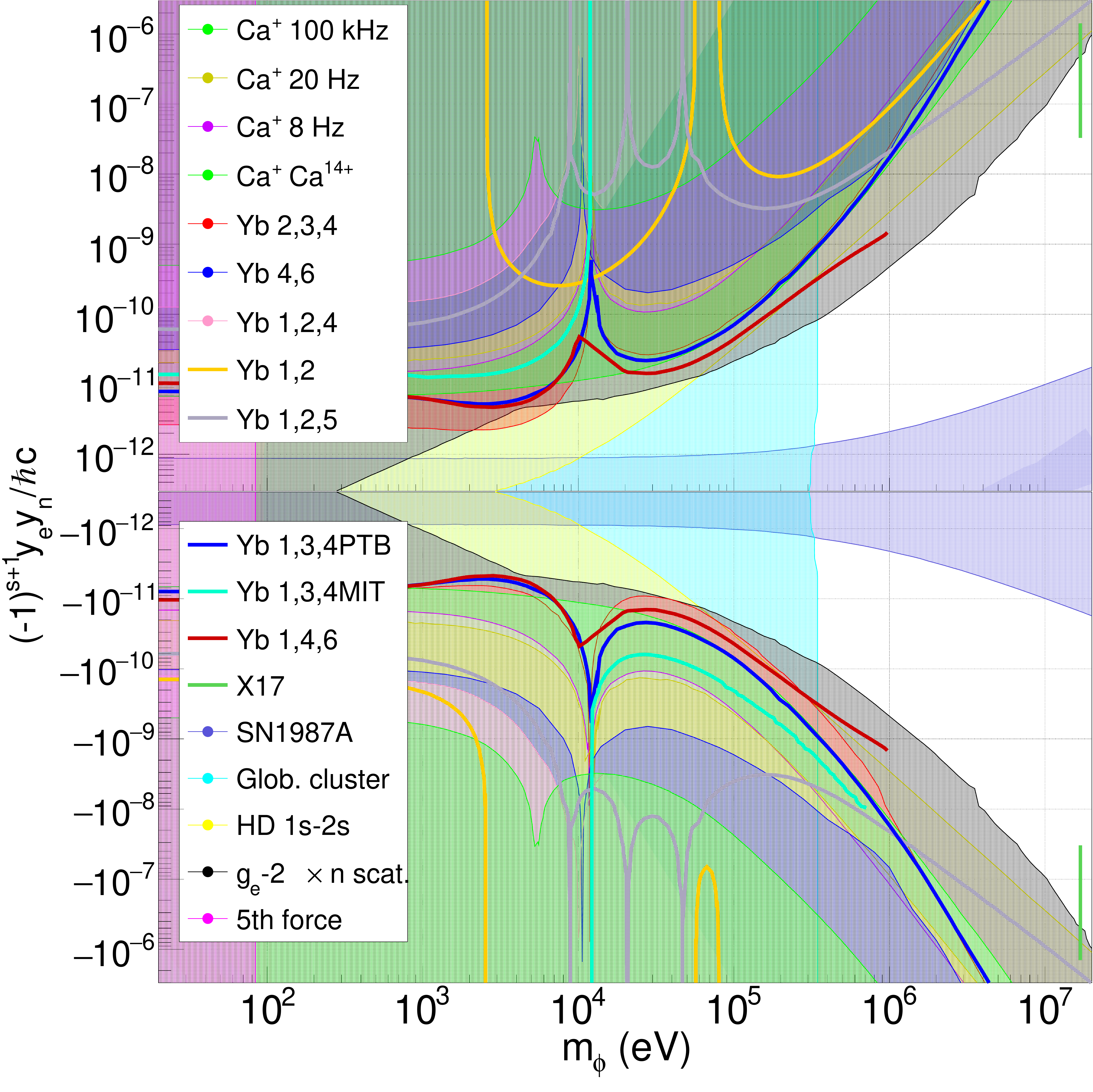}
	\caption{Constraints on a new boson mediating forces between an electron and a neutron: shaded areas are excluded regions, while thick solid lines indicate the best fit for allowed regions. The numbers following Yb in the legend correspond to the following transitions in Yb:  1, 2, and 3 are the $^2$S$_{1/2}\rightarrow^2$D$_{5/2}$, $^2$S$_{1/2}\rightarrow^2$D$_{3/2}$, and $^2$S$_{1/2}\rightarrow^2$F$_{7/2}$ transitions in Yb$^+$ ions, and 4, 5, and 6 are the $6s^2~^1$S$_0\rightarrow6s6p^3$P$_0$, the $6s^2~^1$S$_0\rightarrow5d6s~^1$D$_2$, and $6s^2~^1$S$_0\rightarrow4f^{13}5d6s^2(J=2)$ transitions in neutral Yb, respectively. 
Constraints are cited from 
Ca$^+$ 100 kHz \cite{PhysRevLett.120.091801},
Ca$^+$ 20 Hz \cite{PhysRevLett.125.123003},
Ca$^+$ 8 Hz \cite{PhysRevA.110.L030801},
Ca$^+$ Ca$^{14+}$ \cite{PhysRevLett.134.233002},
Yb $2,3,4$ \cite{PhysRevLett.128.163201},
Yb $4,6$ \cite{PhysRevA.109.062806},
Yb $1,2,4$ \cite{PhysRevX.12.021033},
Yb $1,2$ \cite{PhysRevLett.125.123002},
Yb $1,2,5$ \cite{PhysRevLett.128.073001},
Yb $1,3,4$PTB \cite{PhysRevLett.134.063002},
Yb $1,3,4$MIT \cite{PhysRevLett.128.163201},
Yb $1,4,6$ \cite{2505.04154}. 
X17 is an allowed region associated with the ATOMKI anomaly \cite{PhysRevLett.116.042501,PhysRevLett.134.063002}. 
SN1987A is the constraint due to the energy loss in a supernova \cite{PhysRevD.86.015001,PhysRevLett.134.233002}. 
Glob. cluster  shows the excluded region due to the stellar evolution constraints in the globular clusters \cite{PhysLettB.173.237,PhysRevLett.134.063002}. 
$g_e-2\times$ n scat. \cite{PhysRevLett.134.063002} results from electron $g-2$ measurements \cite{PhysRevLett.130.071801} and neutron scattering \cite{PhysRevD.77.034020}. 
5th force denotes constraints from fifth force searches\cite{PhysRevA.62.052109,PhysRevLett.134.063002}. HD 1s-2s shows the constraints derived from isotope shifts between hydrogen and deuterium in the $1s_{1/2}-2s_{1/2}$ transition. \cite{PhysRevA.108.052825}.
All constraints and allowed regions are plotted directly from the original papers without modifications. Note that the interpretation of the nonlinearity in Refs. \cite{PhysRevLett.134.063002} and \cite{2505.04154} are to put constraint on the coupling constant stronger than the observed nonlinearity by $1.96\sigma$. However, here, for the consistency with other reports, their nonlinearity is sorted as allowed region. 
	}
	\label{ISDM}
\end{figure}


\subsection{Experimental results}
Since the publications of theory papers around 2018, a large number of experimental results has been reported on this topic, ranging from isotope shift measurements alone to direct constraints on the existence of new forces. For certain atomic species, such as Xe \cite{PhysRevLett.131.161803}, Cd \cite{NewJPhys.24.123040}, and Nd \cite{PhysRevA.101.052505}, isotope shifts were reported in the context of the search for new bosons. The constraints on the existence of new bosons are summarized in Fig.~\ref{ISDM}. The first experimental constraints on the new bosons were established through isotope shift measurements for Yb$^+$ \cite{PhysRevLett.125.123002} and Ca$^+$ \cite{PhysRevLett.125.123003}. In these studies, isotope shift measurements of Ca$^+$ with a 20-Hz uncertainty excluded the existence of new bosons based on the observed linearity of the King plot, whereas nonlinearity is observed in the isotope shift measurements of Yb$^+$ with a 300-Hz uncertainty. The initial Yb$^+$ study focused on two E2 transitions, and later measurements included isotope shifts for the E3 transition with an accuracy of 500 Hz\cite{PhysRevLett.128.163201}. Additionally, multiple transitions in neutral Yb were carefully analyzed. At first, isotope shifts for the $6s^2~^1$S$_0\rightarrow5d6s~^1$D$_2$ transition is reported with $<1$ kHz accuracy \cite{PhysRevLett.128.073001} . Then the isotope shifts for the conveitional $6s^2~^1$S$_0\rightarrow6s6p~^3$P$_0$ clock transition are reported \cite{PhysRevX.12.021033}. Because of its high accuracy on the order of 1 Hz, this data are often used for analyses in other reports. The $6s^2~^1$S$_0\rightarrow4f^{13}5d6s^2(J=2)$ transition is first incorporated into a King-plot analysis in Ref. \cite{PhysRevA.109.062806}, with a recent improved report of accuracy \cite{2505.04154}. The most recent results for the E2 and E3 transitions of Yb$^+$ achieved uncertainties of 5 and 16 Hz, respectively \cite{PhysRevLett.134.063002}. While Ca$^+$ and some Yb measurements did not observe nonlinearity, consistent with the constraint based on the $g-2$ measurement for electrons and neutron scattering, certain Yb dataset exhibit nonlinearity, leading to allowed regions for the new bosons.  These allowed and excluded regions contradict each other, suggesting that the observed nonlinearity may be induced by some background effects within the SM. This calls for further investigation. A recent observation of nonlinearity in Ca \cite{PhysRevLett.134.233002} was based on the combination of the precision spectroscopy of Ca$^+$ and Ca$^{14+}$. This study attributed the observed nonlinearity to the second order mass shift and nuclear polarization, setting most stringent constraint on the new bosons.

If the effect of nuclear charge radii does not have to be eliminated using isotope shift data from another transition, the effect of the new boson can be probed through the spectroscopy of a single transition. Such a constraint was obtained from spectroscopy of $1s_{1/2}-2s_{1/2}$ transition in hydrogen and deuterium\cite{PhysRevA.108.052825}, as shown in Fig. \ref{ISDM}. In this case, nuclear charge radii information is derived from spectroscopy of muonic atoms. 

When nonlinearity is observed in the King plot, its source must be carefully analyzed. For this purpose, nonlinearity is decomposed into the following two components: 
$\zeta_{\pm}=d^{168}-d^{170}\pm(d^{172}-d^{174})$, where $d^A$ is the residual of four data points from a linear fit \cite{PhysRevLett.125.123002} (Note that $\lambda_{\pm}$ in Ref. \cite{PhysRevLett.128.163201} is equivalent to $\zeta_{\pm}$). On the $\zeta_+-\zeta_-$ plane, a single source of nonlinearity with varying strength forms a line passing through the origin. When residuals of King plots for different combination of transitions in Yb are plotted on this plane, the data points approximately align along a single line. This suggests that the observed nonlinearity primarily originates from a single source, which can be explained by nuclear density functional theory calculations. These calculations incorporate several factors, including the surface thickness of nuclear density, relativistic corrections, nuclear deformation, and pairing effects \cite{PhysRevLett.128.163201}. The isotope shift measurements are so precise that they can select a nuclear model that accurately describes the experimental results. \cite{PhysRevLett.128.163201,PhysRevLett.134.063002,2505.04154}. 

After accounting for the influence of the nuclear charge distribution, any remaining nonlinearity still can be carefully examined. If this residual nonlinearity is attributed to new bosons, constraints on the existence of new bosons mediating forces between a neutron and an electron can be derived from isotope shift measurements of Yb. Some analyses of currently available data show statistically significant nonlinearity, while others show linear behavior. As a result, different combinations of transitions produce both allowed and excluded regions, as shown in Fig. \ref{ISDM}.

The rapid advancement of King plot analysis has driven the need for precise mass measurements \cite{ChinPhysC.45.030002,ChinPhysC.45.030003}. Because the inverse mass difference plays an important role, highly accurate mass measurements of different isotopes are essential to ensure that mass uncertainties do not limit the search for new bosons. Following the first report on Yb \cite{PhysRevLett.125.123002}, the mass uncertainty of $^{168}$Yb was improved \cite{IntJMassSpectrometry.458.116435}, and recent report on Yb includes even more precise mass measurements \cite{PhysRevLett.134.063002}. For isotope shift measurements with even smaller uncertainties, further improvements in mass measurements are necessary, unless generalized King plot is implemented \cite{PhysRevResearch.2.043444}. To enhance the sensitivity of these searches, the use of radioactive isotopes \cite{PhysRevResearch.1.033113,PhysRevResearch.2.043444} are proposed.

\section{Summary and Outlook}

High precision measurements with quantum states in atoms can detect small energy shifts induced by phenomena originating from nuclear and particle physics. For future advancements, several key aspects are important. One is improvement in measurement precision. Even within the same experimental system, increased measurement precision enhances sensitivity to nuclear and particle physics phenomena in most of the topics covered in this paper. In the past, improvements in clock uncertainty contributed to searches for ultralight dark matter and time variation of fundamental constants. Second, the use of highly sensitive quantum sensors can significantly improve the sensitivity. A notable example is searches for variation of the fine structure constant. Switching from conventional ion clocks to HCI clocks or nuclear clocks would enhance the sensitivity by orders of magnitude. In addition, a solid theoretical understanding of the measurements is crucial. Theoretical calculations of atomic or molecular structures determine the sensitivity of a quantum sensor to the quantity of interest. Examples include the sensitivity coefficient $K$ for the variation of the fine structure constant. Further improvements in the accuracy of these theoretical calculations can refine the determination of parameters in nuclear and particle physics. Moreover, new theoretical proposals can inspire extensive experimental searches for previously unexplored parameter spaces. Historically, searches for the fifth force using isotope shifts and ultralight dark matter expanded following initial theoretical proposals. With the combination of these aspects, further exploration of nuclear and particle physics using atomic clocks as quantum sensors are expected.

\begin{acknowledgments}
The author thanks Takashi Higuchi, Joonseok Hur, Yevgeny Stadnik, and Takumi Kobayashi for helpful discussions and comments on the manuscript. A.K. acknowledges the support of Japan Society for the Promotion of Science KAKENHI Grants No. JP22H01161, and Japan Science and Technology Agency FOREST Grant No. JPMJFR212S. Data sharing is not applicable to this article as no new data were created or analyzed in this study.
\end{acknowledgments}

\bibliography{ApplPhysRev}

\end{document}